\documentclass[aps,prc,twocolumn,floatfix,superscriptaddress,nofootinbib]{revtex4-1}
\usepackage{graphicx}
\usepackage[pdftex,colorlinks,citecolor=blue,bookmarks]{hyperref}
\usepackage{bm}
\usepackage{amsmath}
\usepackage{longtable}
\begin{document}
\title{Variational approximations to exact solutions in shell-model valence spaces: \\ calcium isotopes in the $pf$-shell}
\author{Benjamin Bally} 
\address{Department of Physics and Astronomy, University of North Carolina, Chapel Hill, North Carolina 27516-3255, USA}
\address{Departamento de F\'isica Te\'orica, Universidad Aut\'onoma de Madrid, E-28049 Madrid, Spain}
\address{Instituto de F\'isica Te\'orica UAM-CSIC, E-28049 Madrid, Spain}
\author{Adri\'an S\'anchez-Fern\'andez} 
\address{Departamento de F\'isica Te\'orica, Universidad Aut\'onoma de Madrid, E-28049 Madrid, Spain}
\author{Tom\'as R. Rodr\'iguez} 
\address{Departamento de F\'isica Te\'orica, Universidad Aut\'onoma de Madrid, E-28049 Madrid, Spain}
\address{Centro de Investigaci\'on Avanzada en F\'isica Fundamental-CIAFF-UAM, E-28049 Madrid, Spain}
\begin{abstract}
We study the performance of self-consistent mean-field and beyond-mean-field 
approximations in shell-model valence spaces. In particular, Hartree-Fock-Bogolyubov, particle-number 
variation after projection and projected generator coordinate methods are applied to obtain 
ground-state and excitation energies for even-even and odd-even Calcium isotopes in the $pf$-shell. 
The standard (and non-trivial) KB3G nuclear effective interaction has been used. The comparison with 
the exact solutions ---provided by the full diagonalization of the Hamiltonian--- shows an outstanding agreement 
when particle-number and angular-momentum restorations are performed and both quadrupole and 
neutron-neutron pairing degrees of freedom are explicitly explored as collective coordinates. 
\end{abstract}
\maketitle
\section{Introduction}
The study of the structure of the atomic nuclei from a microscopic point of view is a very complicated task mainly due to two 
interrelated problems. First, the interaction among the composite nucleons is rather intricate and for a long time could not be  reliably derived from first principles. The recent advent of Effective Field Theories in nuclear physics \cite{Epelbaum09a,Hammer19a} has considerably improved the situation. The other complication arises from the number of particles in a nucleus that is usually too large to solve exactly the problem but not large enough to describe the system using statistical
mechanics. Therefore, one of the keys to the theoretical description of the atomic nucleus is the design and
implementation of efficient quantum many-body techniques.

Over the last decade, \textit{ab initio} methods, which aim at solving as exactly as possible, and in a systemically improvable manner, the many-body Schr\"odinger equation, have made great progress \cite{Soma13a,Hagen14a,Hergert16a,Tichai18a}. In certain specific cases, they can even reach nuclei with $A \sim 100$ nucleons \cite{Morris18a}. 
Their accuracy in the description of experimental data, however, is not yet on par with what is obtained by phenomenological models such as the Interacting Shell Model (ISM) \cite{Caurier05a,Otsuka01a} or the Energy Density Functional (EDF) method \cite{Bender03a,Niksic11a,Egido16a,Robledo18a}.

Within the ISM framework, the problem with $A$ nucleons is reduced to a system with $A'=A-A_{core}$ particles that are interacting in a 
restricted valence space. The resulting Hamiltonian matrix is then directly  diagonalized to obtain the exact solutions of the reduced problem.
The valence space is normally made of a few one-body spherical harmonic oscillator orbits and $A_{core}$ is 
normally a given double-magic nucleus that defines an inert core. 
To this day, the ISM is the most successful method to compute low-lying 
nuclear spectra of nuclei not very far away from shell closures. 

On the other hand, EDF methods consider all the particles in the system but the many-body problem is solved through a variational approximation. The most advanced EDF methods, the so-called Multi Reference EDF (MREDF) approaches, combine the Hartree-Fock-Bogolyubov (HFB) methodology with symmetry restoration and configuration mixing techniques \cite{Duguet14b}. Calculations using Skyrme, Gogny\footnote{In Gogny implementations, MREDF approaches are also known as symmetry conserving configuration mixing (SCCM) methods~\cite{Egido16a,Robledo18a}} and covariant functionals have been widely applied to describe various nuclear 
observables in the whole nuclear chart \cite{Bender03a,Niksic11a,Egido16a,Robledo18a}.

The broadness of the applicability of the variational approaches is one of the main positive aspects of these methods. However, we 
cannot judge in general whether the results provided by these methods are close or far away from the exact solutions. This 
indetermination leads to an uncertainty in the theoretical results that cannot be easily assessed. In this work, we benchmark 
several variational approximation schemes 
against the exact solutions using a shell-model valence space and Hamiltonian as a non-trivial playground. In 
particular, we study the even and odd calcium isotopes in the $pf$-shell using the standard KB3G nuclear interaction. The different 
approaches cover from the simplest HFB method to the most sophisticated projected generator coordinate method (PGCM) that includes particle-number and angular-momentum restorations as well as the configuration mixing along 
the quadrupole and pairing degrees of freedom. 

Similar implementations to the present approach have been recently used to evaluate neutrinoless double-beta decay nuclear matrix 
elements~\cite{Jiao17a,Yao18a,Jiao18a}. Some of them consider a limited number of 
collective degrees of freedom as generating coordinates and none of them implements the particle-number variation after projection 
(PNVAP) method. Additionally, only even-even nuclei were computed so far with valence-space Hamiltonian-based GCM. 
A study of HFB and PNVAP potential energy surfaces along the axial quadrupole degree of freedom in $sd$- and $pf$-shell even-even 
nuclei has been also reported in Ref.~\cite{Maqbool11a} but neither angular momentum restoration nor configuration mixing 
were performed in that work. Moreover, Hartree-Fock (HF) total energy surfaces along quadrupole degrees of freedom have been used to interpret  
ISM~\cite{Hadynska16a,BounthongPHD} and Monte Carlo Shell Model 
(MCSM)~\cite{Tsunoda14a,Togashi18a} calculations.
prominently
Also, simpler Hamiltonians like Pairing-plus-quadrupole 
(PPQ) interactions have been also used in the past within the projected shell model (PSM) 
framework~\cite{Hara95a,Sun16a} and with GCM approximations including two 
quasi-particle excitations explicitly~\cite{Chen16a,Chen17a}. 
Our goal is to go beyond these 
simplistic Hamiltonians although, for the moment, some single-particle degrees of freedom included in the above methods will be omitted.
Finally, we note that the idea to build sophisticated variational approximations is at the core of the famous variation after mean-field projection in realistic model spaces (VAMPIR) \cite{Schmid87a,Schmid04a} approach that includes the variation after projection (VAP) onto particle number and angular momentum simultaneously. The angular-momentum variation after projection although highly desirable is numerically very intensive in the general case, i.e.\@ considering many states without any symmetry restrictions and projected onto other quantum numbers, and therefore is not considered here.

The paper is organized as follows. First, in Sec.~\ref{Theoretical_Framework}, 
we define the valence space, the  Hamiltonian, and the different variational approaches used in this work. Then, in Sec.~\ref{Results}, we show how these methods are implemented using the nucleus $^{48}$Ca as an 
example. From there, we extend the study to the rest of even and odd calcium 
isotopes. Finally, a summary and outlook is given in Sec.~\ref{Summary}.
\section{Theoretical framework}
\label{Theoretical_Framework}
\subsection{Basic principles} 
Let us first consider a model space spanned by a set of harmonic oscillator single-particle states $\left\{ \phi_a \right\}$, associated with the set of creation and annihilation operators $\left\{ c_a ; c^{\dagger}_a \right\}$, which are characterized by their principal quantum number $ n_a$, orbital angular momentum $l_a$, spin $s_a = 1/2$, total angular momtum $j_a$ and its third component $m_{j_a}$, and isospin $t_a=1/2$ and its third component $m_{t_a}$.\footnote{We use as a convention $m_{t_a} = -1/2$ for proton single-particle states and $m_{t_a} = +1/2$ for neutron single-particle states.} 
For the sake of clarity, we use the shorthand notation $a \equiv (n_a , l_a, s_a , j_a, m_{j_a}, t_a, m_{t_a} )$.

Given this model space, the Hamiltonian\footnote{The inclusion of three- or higher-body forces would not pose any conceptual problem.} of the system reads
\begin{equation}
\hat{H} = \sum_{ab} t_{ab} c^{\dagger}_{a} c_{b} + \frac{1}{(2!)^2} \sum_{abcd} \bar{v}_{abcd} c^{\dagger}_{a} c^{\dagger}_{b} c_{d} c_{c} , 
\end{equation}
where $t_{ab}$ are one-body matrix elements and $\bar{v}_{abcd}$ are (antisymmetrized) two-body matrix elements.

In the present work, the model space under consideration is restricted to the valence space made of the $pf$-shell
spherical orbits ($0f_{7/2}$, $1p_{3/2}$, $1p_{1/2}$ and $0f_{5/2}$). As for the Hamiltonian, we use the matrix elements of the KB3G interaction~\cite{Poves01a}, which in particular provides us with the single-particle energies $t_{ab} = \epsilon_a \delta_{ab}$.

In such a small model space, it is possible to solve exactly the many-body Schr\"{o}dinger equation for any given numbers of valence protons, $Z$, and neutrons, $N$, by performing a direct diagonalization of the Hamiltonian matrix in the many-body basis made of all $A'$-particles Slater determinant, with $A'=Z+N$. This is the standard ISM framework and, in this work, the ISM solutions are obtained using the code ANTOINE developed by the Strasbourg-Madrid collaboration~\cite{Caurier05a}.

On the other hand, the number of configurations and, correspondingly, the dimensions of the Hamiltonian matrices increase 
combinatorially with the number of active particles and the size of the valence space. This fact forbids the calculation of exact solutions 
in the general case and, therefore, approximate methods are in order. One of the most widely used is the PGCM \cite{RS80a,Bender03a,Robledo18a} within which the nuclear many-body wave functions are taken to be linear combinations of the form
\begin{equation}
|\Psi^{JMNZ\pi}_{\sigma} \rangle=\sum_{qK} f^{JMNZ\pi}_{\sigma;qK} P^{J}_{MK} P^{N} P^{Z} P^{\pi} |\Phi(q)\rangle ,
\label{GCM_wf1}
\end{equation}
where $\left\{|\Phi(q)\rangle\right\}$ are the so-called \emph{intrinsic} states, assumed here to be Bogolyubov quasi-particle states, 
$P^{N(Z)}$ is the projector onto good number of neutrons (protons), 
$P^{J}_{MK}$ is the angular momentum projection operator, $P^{\pi}$ is the parity projector. The precise form of the projection operators can be found somewhere else \cite{Bender03a,RS80a,BallyPHD}. For the sake of clarity, we regroup the set of quantum numbers under the label $\Gamma \equiv (JMNZ\pi)$.
The coefficients $f^{\Gamma}_{\sigma;qK}$ are 
obtained by solving the Hill-Wheeler-Griffin (HWG) equations~\cite{RS80a}
\begin{equation}
\sum_{q'K'} \left( \mathcal{H}^{\Gamma}_{qK,q'K'} - E^{\Gamma}_{\sigma} \mathcal{N}^{\Gamma}_{qK,q'K'} \right) f^{\Gamma}_{\sigma;q'K'} = 0 ,
\label{HWG_1}
\end{equation}
where we have defined the Hamiltonian and norm overlaps matrices
\begin{subequations}
\begin{align}
\mathcal{H}^{\Gamma}_{qK,q'K'}&= \langle \Phi(q)|\hat{H}P^{J}_{KK'}P^{N}P^{Z}P^{\pi}|\Phi(q')\rangle , \\
\mathcal{N}^{\Gamma}_{qK;q'K'}&= \langle \Phi(q)|P^{J}_{KK'}P^{N}P^{Z}P^{\pi}|\Phi(q')\rangle .
\end{align}
\end{subequations}
The associated generalized eigenvalue problem can be transformed into a normal eigenvalue problem by removing the linear dependencies of 
the original set of projected states $\left\{  P^{J}_{MK} P^{N} P^{Z} P^{\pi} |\Phi(q) \rangle \right\}$. Hence, the nuclear states given in 
Eq.~\eqref{GCM_wf1} are expressed in the so-called \emph{natural basis} that is obtained by finding first the eigenvalues and eigenvectors of the 
norm overlap matrix
\begin{equation}
\sum_{q'K'} \mathcal{N}^{\Gamma}_{qK,q'K'}u^{\Gamma}_{\lambda;q'K'} = n^{\Gamma}_{\lambda} u^{\Gamma}_{\lambda;qK} , 
\end{equation}
then, the orthonormal states are defined as
\begin{equation}
|\Lambda^{\Gamma}_{\lambda} \rangle = \sum_{qK} \frac{u^{\Gamma}_{\lambda;qK}}{\sqrt{n^{\Gamma}_{\lambda}}} P^{J}_{MK} P^{N} P^{Z} P^{\pi} |\Phi(q)\rangle ,
\label{nat_basis}
\end{equation}
where we set a threshold for the smallest norm overlap matrix eigenvalue, $n^{\Gamma}_{\lambda}>\epsilon$. 
The nuclear states, Eq.~\eqref{GCM_wf1}, can now  be written as
\begin{equation}
|\Psi^{\Gamma}_{\sigma} \rangle = \sum_{\lambda} G^{\Gamma}_{\sigma;\lambda}|\Lambda^{\Gamma}_{\lambda}\rangle , 
\label{GCM_wf2}
\end{equation}
and the HWG equations 
\begin{equation}
\sum_{\lambda'}\langle\Lambda^{\Gamma}_{\lambda}|\hat{H}|\Lambda^{\Gamma}_{\lambda'}\rangle G^{\Gamma}_{\sigma;\lambda'} = E^{\Gamma}_{\sigma} G^{\Gamma}_{\sigma;\lambda} .
\end{equation}
Finally, the energies $E^{\Gamma}_{\sigma}$ computed with the PGCM method are approximations to the exact eigenvalues of $\hat{H}$. 
In addition, the coefficients $G^{\Gamma}_{\sigma;\lambda}$ can be used to evaluate the collective wave functions, 
occupation numbers, transition probabilities, electromagnetic moments, etc.~\cite{Bender03a,Robledo18a}. 

We note in passing that our Hamiltonian-based method is free from the problems of self-interaction and self-pairing appearing in MR EDF calculations \cite{Anguiano01a,Lacroix09a,Bender09a,Duguet09a}, which in particular implies that all the direct, exchange and pairing terms are included. Also, to compute the norm overlap between two quasi-particle vacua, we use the Pfaffian method \cite{Robledo09a,Avez12a} with the numerical routines published in Ref.~\cite{Wimmer12a}. Other methods have been recently proposed \cite{Bally18a,Mizusaki18a}.
\subsection{Generation of the intrinsic states} 
The quality of the PGCM approach relies heavily on the richness of the underlying set of selected intrinsic wave functions $\left\{|\Phi(q)\rangle\right\}$.
In the present case, all of them are Bogolyubov quasi-particle wave functions, i.e.\@ they are vacua for a set of quasi-particle operators $\left\{ \beta_a (q) ; \beta^{\dagger}_a (q) \right\}$
defined through unitary linear Bogolyubov transformations
\begin{equation}
\begin{pmatrix}
\beta (q)\\
\beta^{\dagger}(q)
\end{pmatrix}
= {\cal W}^{\dagger}(q) 
\begin{pmatrix}  
c \\
c^{\dagger} 
\end{pmatrix}  \, , \label{bogophi} \\
\end{equation}
where
\begin{equation}
{\cal W}(q) \equiv
\begin{pmatrix}
U(q) & V^{\ast}(q) \\
V(q) &  U^{\ast}(q)
\end{pmatrix} .
\end{equation} 
The matrices $U(q)$ and $V(q)$ are variational parameters that are obtained by solving either HFB or particle-number variation after projection (PNVAP) equations (see below). There are several aspects that characterize a set of intrinsic states. First, the above transformations can be 
restricted and the quasi-particles would fulfill certain self-consistent symmetries. In the present work, we allow for general HFB 
transformations without any restriction except
for the fact that the matrices $U(q)$ and $V(q)$ are kept real.
In particular, angular-momentum, particle-number and 
parity symmetry breaking, as well as proton-neutron mixing are allowed. The latter is key to include proton-neutron pairing in the wave functions and it is normally ignored in most of the EDF calculations. However, it will be irrelevant in the study of the calcium isotopic 
chain within the $pf$-shell since only neutrons are involved in the valence space. Moreover, all the single-particle orbits in the $pf$-shell 
have negative parity and, therefore, only positive (negative) parity states can be obtained for even (odd) nuclei. 
As a consequence, parity is not broken by the HFB transformations and the parity projection is not needed.

Second, we distinguish between HFB and VAP approaches~\cite{RS80a} depending on 
the definition of the energy minimized in the variational process. In this study, we perform either HFB or PNVAP calculations. The PNVAP method is variationally better for ground state energies, provides larger pairing 
correlations and is free from spurious phase transitions in the pairing channel that occur in HFB 
approaches~\cite{Anguiano01a,Rodriguez05a,Rodriguez07a}. Nevertheless, we will 
compare the results obtained by PGCM schemes using both HFB and PNVAP underlying states. In both cases, the gradient 
method \cite{Robledo11a,Egido95a} complemented by a momentum term \cite{Ryssens19a} is used to solve the non-linear equations at hand.

The intrinsic wave functions of nuclei with an odd number of particles are obtained by the blocking method to ensure their correct (odd) 
number-parity symmetry~\cite{RS80a}. The solution of the variational equations could depend on the initial blocked state 
(the seed wave function needed to start the calculation) \cite{Bally14a,BallyPHD}. This is particularly important in HFB calculations where local minima 
that correspond to ground or excited states could be obtained by the gradient method. In the present work, we try several blocked initial 
configurations. Then, we take the configuration that gives the lowest energy after convergence as the actual solution.  
On the other hand, PNVAP calculations are much less dependent on the choice of the seed state and the gradient method leads normally to 
the absolute minimum. 
The choice to select only the lowest self-consistent quasi-particle state for given set of constraint may warrant, \emph{a priori}, a good variational exploration of excitated states.
But as we will see later on, we obtain an excellent agreement with the exact results also for those states.

The third important aspect about the definition of the set of intrinsic states is the selection of the collective coordinates, $q$. 
Hence, the variational equations are solved with constraints, i.e., the energy that is minimized is modified 
with Lagrange multipliers. These terms ensure that the conditions $\langle \Phi(q)|\hat{Q}|\Phi(q)\rangle=q$ are fulfilled ($\hat{Q}$ are the 
operators associated to the collective coordinates). The most important collective degrees of freedom present in realistic nuclear 
interactions are the lowest multipole deformations, 
particularly the quadrupole deformations, and pairing~\cite{Dufour96a}. The former can be explored by imposing constraints 
in the values of the mass quadrupole operators
\begin{subequations}
\begin{align}
\hat{Q}_{2\mu} &= r^{2} Y_{2\mu}(\theta,\varphi) , \\
q_{20} &= \langle\Phi(q)|\hat{Q}_{20}|\Phi(q)\rangle ,\\
q_{21} &= \frac12 \langle\Phi(q)|\hat{Q}_{21}-\hat{Q}_{2-1}|\Phi(q)\rangle , \\
q_{22} &= \frac12 \langle\Phi(q)|\hat{Q}_{22}+\hat{Q}_{2-2}|\Phi(q)\rangle .
\end{align}
In the present calculations, we always set $q_{21}=0$. In such a case, we can also define the parameters $\beta_{2}$ and $\gamma$ as
\begin{align}
q_{20} &= C \beta_{2} \cos\gamma , \\
q_{22} &= \sqrt{2} C \beta_{2} \sin\gamma ,
\end{align}
\end{subequations}
where $C=\frac{3r_{0}^{2}A^{5/3}}{8\pi}$, $r_{0}=1.2$ fm and $A$ is the total mass number (including core and valence space particles). 
Additionally, electromagnetic transitions and moments are calculated with the effective charge $e_{p}$ for protons and $e_{n}$ for neutrons.
In the $pf$-shell, we choose the standard values $1.5$ and $0.5$ for protons and neutrons, 
respectively~\cite{Caurier05a,Otsuka01a}. 

Pairing degrees of freedom can be explored by constraining the expectation value of a pair creation operator with respect to the intrinsic states. In this work, we use an operator that couples pairs within a given orbit $\breve{a} \equiv  (n_a, l_a , j_a , s_a , t_a)$ to a good total angular momentum $J$ and total isospin $T$ \cite{Talmi93a,Hinohara14a}
\begin{equation}
\begin{split}
\left[\hat{P}^{\dagger}\right]^{JT}_{M_{J}M_{T}} &= \sum_{\breve{a}} \left[\hat{P}^{\dagger}_{\breve{a}}\right]^{JT}_{M_{J}M_{T}} \\
&= \frac{1}{\sqrt{2}}\sum_{\breve{a}}\sqrt{2j_{a}+1}\left[c^{\dagger}_{\breve{a}}c^{\dagger}_{\breve{a}}\right]^{JT}_{M_{J}M_{T}}
\end{split}
\end{equation}
where the creation operators are $JT$-coupled according to 
\begin{equation}
\begin{split}
\left[c^{\dagger}_{\breve{a}}c^{\dagger}_{\breve{b}}\right]^{JT}_{M_{J}M_{T}} 
= &\frac{1}{\sqrt{1+\delta_{j_{a}j_{b}}}}\sum_{\substack{m_{j_{a}}m_{j_{b}} \\ m_{t_{a}}m_{t_{b}}}}\langle j_{a}m_{j_{a}}j_{b}m_{j_{b}}|JM_{J}\rangle \\
& \times \langle 
\tfrac{1}{2}m_{t_{a}}\tfrac{1}{2}m_{t_{b}}|TM_{T}\rangle c^{\dagger}_{a}c^{\dagger}_{b} .
\end{split}
\end{equation}
Both isoscalar ($T=0$, $J=1$) $pn$-pairing and isovector ($T=1$, $J=0$) $pp$-, $nn$-, and $pn$-pairing can be explored with these operators. 
In this work, we only study the $nn$-pairing channel ($T=1$, $M_{T}=1$, $J=0$, $M_{J}=0$) because only neutrons are present in the calcium 
chain in the $pf$-shell, i.e., the intrinsic wave functions can be constrained to
\begin{equation}
\delta_{nn}= \frac12 \langle \Phi(q) | \left[\hat{P}\right]^{01}_{01} +
\left[\hat{P}^{\dagger}\right]^{01}_{01} | \Phi(q) \rangle .
\label{nn_pair}
\end{equation}

In Table~\ref{table1}, we summarize and label the different PGCM schemes that are examined in the present work depending on 
the type of energy minimization scheme used (HFB or PNVAP) and the collective coordinates explored. In all cases, particle-number (proton and neutron) and angular-momentum (three Euler angles) 
projections were performed. The number of integration points
taken to discretize the integrals over the gauge and Euler angles was large enough as to ensure a full convergence in the nominal expectation values of the particle-number and angular-momentum operators computed with the GCM wave functions (Eq.~\eqref{GCM_wf1}). The PGCM calculations were performed using the newly developed software TAURUS~\cite{Bally19a}.
\begin{table}
\caption{Different PGCM approximation schemes used to compute the structure of calciums isotopes.}
\begin{ruledtabular}
\begin{tabular}{ccc}
Label & Energy minimization & Collective coordinates \\
\hline
PGCM$_{1}$ & HFB & $(\beta_{2},\gamma)$ \\
PGCM$_{2}$ & PNVAP & $(\beta_{2},\gamma)$ \\
PGCM$_{3}$ & PNVAP & $(\beta_{2},\gamma,\delta_{nn})$
\end{tabular}
\end{ruledtabular}
\label{table1}
\end{table}%
\section{Results}
\label{Results}
\subsection{${}^{48}$Ca}
In this section, we illustrate our methodology taking the nucleus $^{48}$Ca as an example. It is noteworthy that, as it will be demonstrated below, it represents one of the most difficult cases for our model. 

As mentioned in the previous section, the first step in our method is the construction of a set of quasi-particle states through a series of constrained HFB/PNVAP calculations. It is important to point out that in a restricted valence space the range of admissible values for the constraints is much more limited than in a no-core implementation. 
Indeed, working with a handful of particles and single-particle states, it is not possible to build a many-body wave function that satisfies any arbitrary values of the constraints. For example, the largest $\beta_2$ value reachable in the model space is quite small compared to the values used in traditional EDF calculations. In the present work, the bounds of the constraints are determined heuristically.
\begin{figure}[tb]
\begin{center}
\includegraphics[width=\columnwidth]{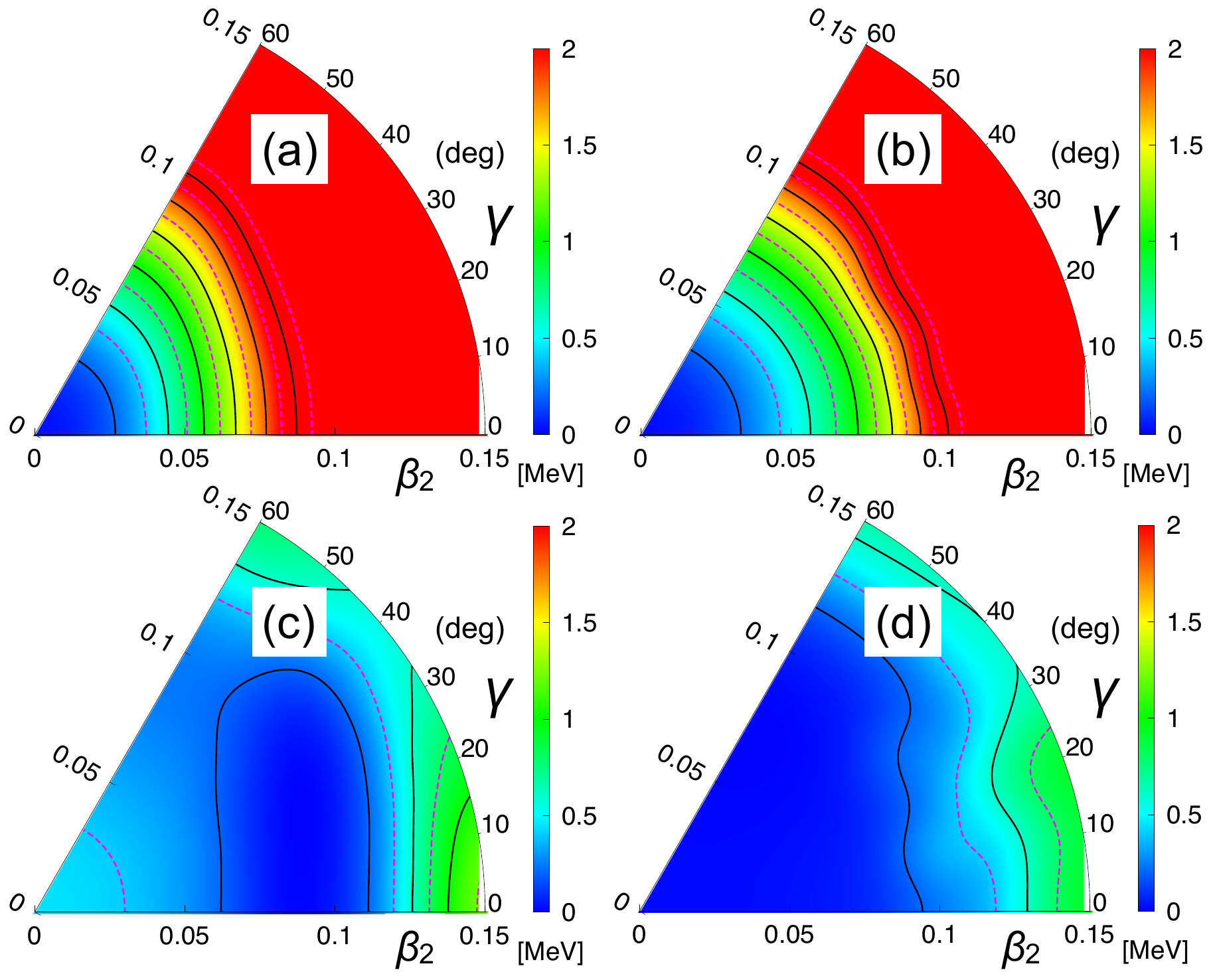}
\end{center}
\caption{(color online) Total energy surfaces (TES) as a function of the quadrupole degrees of 
freedom, $(\beta_{2},\gamma)$ calculated for the nucleus $^{48}$Ca using the following approaches: 
(a) HFB, (b) PNVAP, 
and their corresponding particle-number and angular momentum projected (PNAMP, $J=0$) total energy surfaces, 
(c) HFB+PNAMP and (d) PNVAP+PNAMP. Surfaces are normalized to their respective minimum, i.e., 
(a) -6.446 MeV, (b) -7.195 MeV, (c) -6.896 MeV and (d) -7.209 MeV. Contour lines are separated by 0.2 MeV.}
\label{Fig1}
\end{figure}

The total energy as a function of the quadrupole degrees of freedom, $(\beta_{2},\gamma)$, is represented in Fig.~\ref{Fig1}. On the top panels, the HFB 
(Fig.~\ref{Fig1}(a)) and PNVAP (Fig.~\ref{Fig1}(b)) total energy surfaces (TES) are shown. As expected in this doubly-magic nucleus, in 
both cases the absolute minimum is located at the spherical configuration and the energy rises quickly with $\beta_{2}$ and is almost 
independent of $\gamma$; also we observe that the PNVAP surface is slightly softer. At the spherical point, the pairing collapses in the HFB calculation, which is thus reduced to a plain HF minimization. By contrast, including pairing correlations through the PNVAP permits to gain about 750 keV extra binding energy. However, the main differences between the two approaches are found whenever the particle-number and angular 
momentum projection ($J=0$) are performed on their respective intrinsic states. In the PNVAP+PNAMP case (Fig.~\ref{Fig1}(d)), we obtain an 
interval in $\beta_{2}\in[0.0,0.1]$ where the TES is almost flat independently of the $\gamma$ degree of freedom. The energy gain is 
small with respect to the spherical minimum obtained previously. On the other hand, the HFB+PNAMP results also show some $
\gamma$-softness but the absolute minimum is found at larger prolate deformations. Moreover, the energy gain is larger (740 keV) 
due to the particle-number projection that is not present in the HFB approach. Nevertheless, such an energy is still 300 keV above the 
energy obtained by a PNVAP+PNAMP method. The exact ground state provided by the ISM diagonalization $(E(0^{+}_{1})=-7.569$ 
 MeV) is still below these minima (see caption of Fig.~\ref{Fig1}).

\begin{figure}[tb]
\begin{center}
\includegraphics[width=0.65\columnwidth]{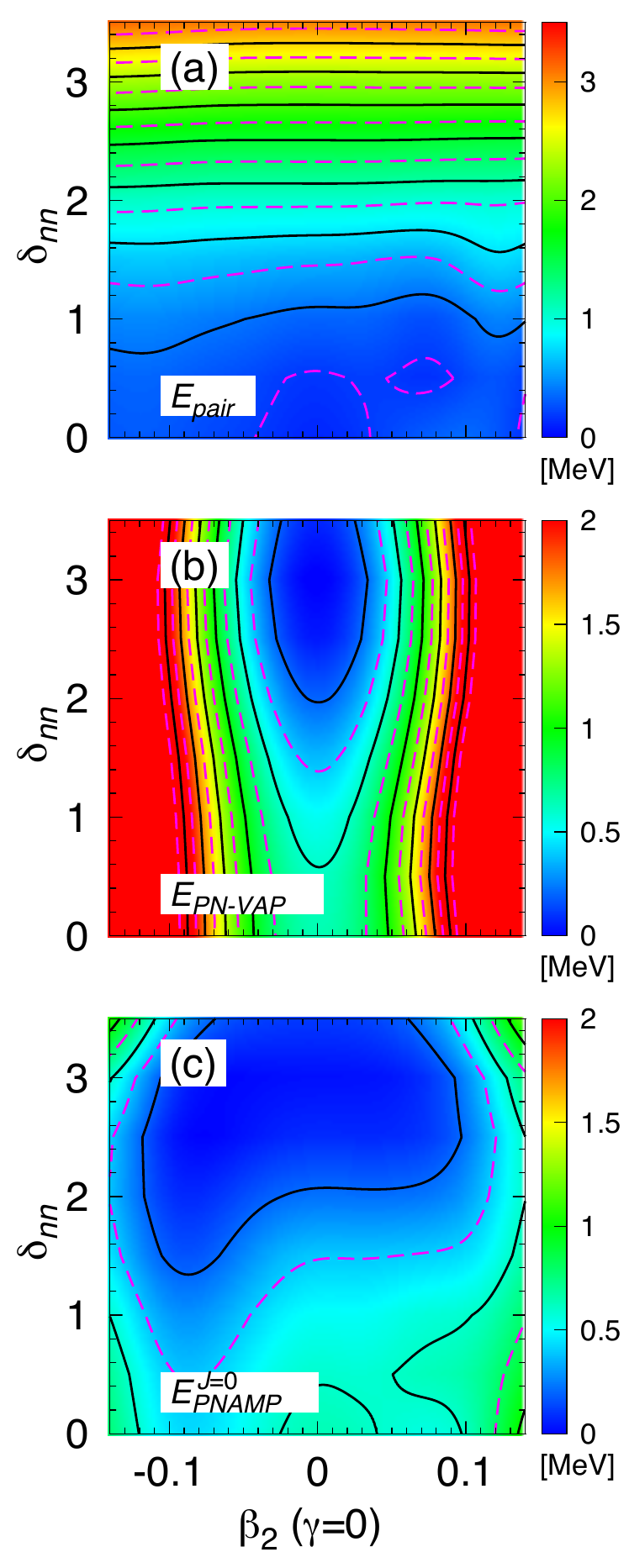}
\end{center}
\caption{(color online) (a) Intrinsic pairing energy, (b) particle-number projected and (c) particle-number and angular-momentum 
projected $(J=0)$ total energy surfaces 
as a function of the axial quadrupole  $(\beta_{2},\gamma=0^\circ \text{ or } 180^\circ)$ and $nn$-pairing $(\delta_{nn})$ degrees of freedom calculated 
for the nucleus $^{48}$Ca using the PNVAP intrinsic wave functions. The sign of the pairing energy has been 
inverted and the total energy surfaces are normalized to their minima (-7.195 MeV and -7.221 MeV, respectively). 
Contour lines are separated by 0.2 MeV.}
\label{Fig2}
\end{figure}
We now study the role of the $nn$-pairing correlations. To better visualize the influence of this degree of freedom, 
we display in Fig.~\ref{Fig2} the results performing PNVAP calculations with constraints on the axial 
quadrupole deformation $(\beta_{2},\gamma=0^{\circ} \text{ or } 180^{\circ})$ and the isovector pairing content $(\delta_{nn})$ defined in 
Eq.~\eqref{nn_pair}. In Figure \ref{Fig2}(a), we show the intrinsic $nn$-pairing energy~\cite{RS80a}, 
$E_{pair}=-\tfrac{1}{2}\mathrm{Tr}\left(\Delta\kappa^{*}\right)$, calculated using the PNVAP intrinsic wave functions without any 
projections.\footnote{i.e.\@ using the pairing density $\kappa_{ab} = \langle \Phi |  c_b c_a | \Phi \rangle$ and the pairing field $\Delta_{ab} = \frac12 \sum_{cd} \bar{v}_{abcd} \kappa_{cd}$. } 
We observe that the pairing energy grows monotonically with 
increasing values of $\delta_{nn}$ and is also almost independent of $\beta_{2}$ (contour lines are horizontal). Therefore,
the effect of varying the parameter $\delta_{nn}$ is just a way of changing the pairing content of the intrinsic wave functions~\cite{Lopez11a}. 
Concerning the total energies (Figs.~\ref{Fig2}(b)-(c)), we observe a similar behavior to the $(\beta_{2},\gamma)$ 
case when we compare the PNVAP
and the angular momentum projected $(J=0)$ surfaces. In the former, a distinct minimum around 
the spherical point $(\beta_{2}=0)$ and 
$\delta_{nn}\in[1.5,3.5]$\ is found. This surface is softer in the pairing direction than in the quadrupole direction. The angular
momentum projection, however, produces a much flatter TES (Fig.~\ref{Fig2}(c)) although slightly lower energies are obtained for 
larger values of $\delta_{nn}$. The energy difference between the minimum before and after projection is rather small ($\sim26$ keV). This behavior, together with the degeneracy also found in the angular momentum projected surface 
in the $(\beta_{2},\gamma)$ plane, suggests that all these degrees of freedom should be taken into account simultaneously in a 
GCM calculation. 

\begin{figure}[tb]
\begin{center}
\includegraphics[width=\columnwidth]{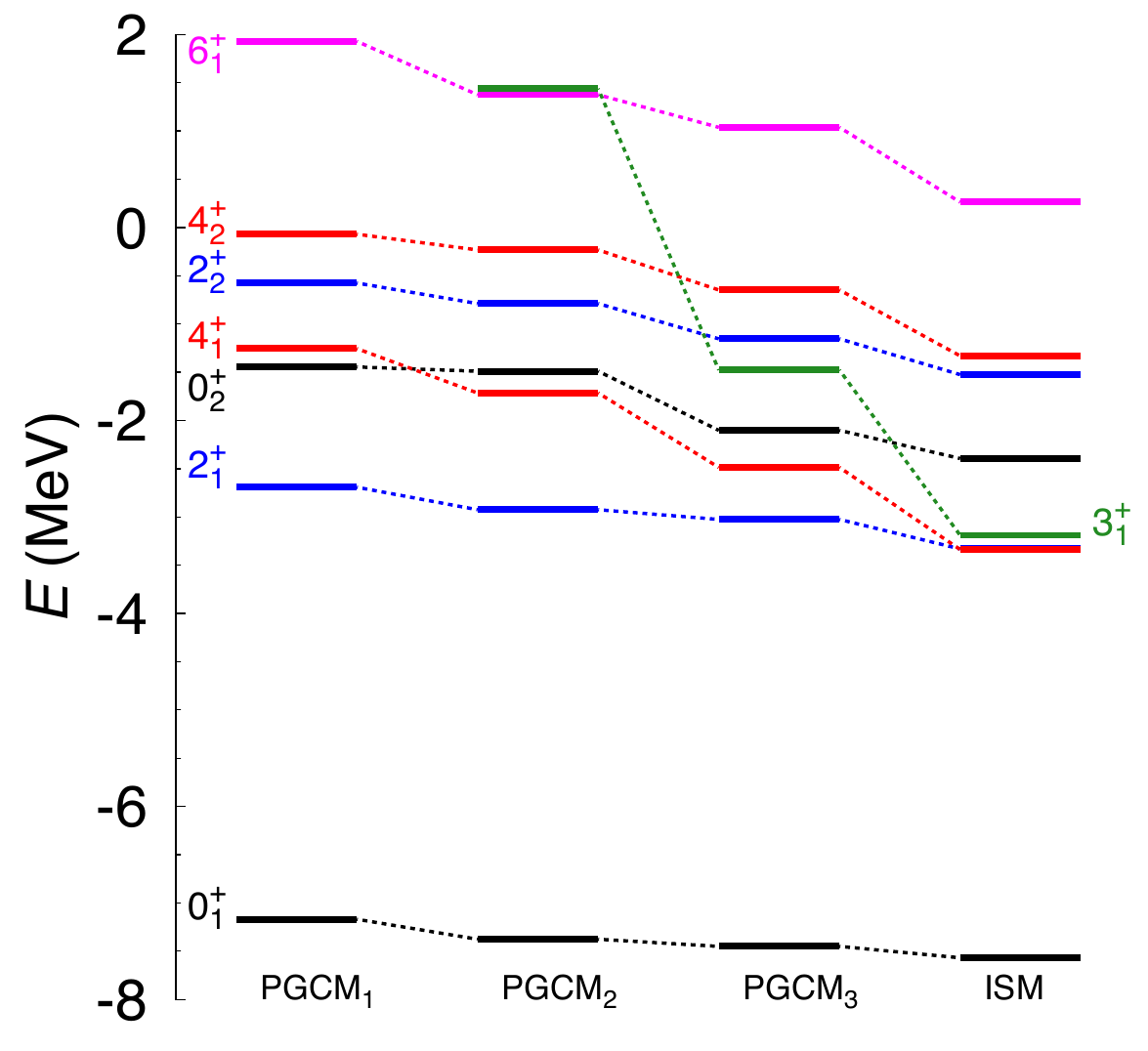}
\end{center}
\caption{(color online) Energy spectrum for $^{48}$Ca calculated exactly (within the ISM framework) and 
with PGCM methods using the KB3G interaction in the $pf$-shell.}
\label{Fig3}
\end{figure}
We compare now the different PGCM calculations (see Table~\ref{table1}) performed for the nucleus $^{48}$Ca with the exact results provided
by the ISM framework. In Figure \ref{Fig3}, the lowest eigenvalues for angular momenta $J=$ 0, 2, 3, 4 and 6 are shown.
We first notice that all of the PGCM approaches are variational approximations to the exact solution plotted on the right, i.e., the energies are
all above the energies obtained from the diagonalization of the Hamiltonian. We clearly see an improvement of the results whenever
PNVAP rather than HFB intrinsic wave functions are used (from PGCM$_{1}$ to PGCM$_{2}$). Additionally, the explicit exploration of
the pairing correlations (PGCM$_{3}$) provides even closer results to the exact solutions. In fact, the difference between the exact and 
PGCM$_{3}$ ground-state energies is $119$ keV showing the ability of these methods to approach the exact values. The 
 reproduction of excited states with a larger value of angular momentum, however, is not as good as for the ground state. In particular, we found that the $3^{+}$ 
state is particularly sensitive to the variational approach. This effect can be understood as our minimization schemes probe primarily the ground-state energy. 
Such a deficiency can be corrected 
within the PGCM framework by enlarging the initial set of quasi-particle wave functions with: a) intrinsically rotating states through cranking 
calculations~\cite{Borrajo15a,Egido16b}; and/or, b) multi-quasi-particle excitations obtained through the blocking 
mechanism~\cite{Chen16a,Chen17a}. Both improvements will be explored in the future but are beyond 
the scope of the present work.

Electric quadrupole transitions and moments are also computed and the results of the most relevant ones are written in 
Table~\ref{48Ca_BE}. We observe a slight overestimation of the exact results in most of the transitions and moments. Furthermore, 
the PGCM$_{3}$ method shows, as expected, the best agreement, and lower values of the spectroscopic quadrupole
moments are obtained whenever the pairing correlations are better described (PNVAP vs.\@ HFB minimization). 
\begin{table}
\caption{Reduced transition probabilities, $B(E2)$, and electric spectroscopic quadrupole moments, $Q$, calculated with different 
methods using proton (1.5) and neutron (0.5) effective charges. The $B(E2)$ are given in units of $e^{2}$fm$^{4}$ whereas the $Q$ are expressed in units of $e$fm$^{2}$.}
\begin{ruledtabular}
\begin{tabular}{ccccc}
 & PGCM$_{1}$ & PGCM$_{2}$ & PGCM$_{3}$ & ISM \\
\hline
$B(E2:2^{+}_{1}\rightarrow0^{+}_{1})$ & 12.7 & 12.7 & 12.7 & 11.5 \\
$B(E2:2^{+}_{1}\rightarrow0^{+}_{2})$ & 0.8 & 0.9 & 1.0 & 1.0 \\
$B(E2:2^{+}_{2}\rightarrow0^{+}_{1})$ & 0.0 & 0.0 & 0.0 & 0.0 \\
$B(E2:2^{+}_{2}\rightarrow0^{+}_{2})$ & 30.4 & 24.5 & 23.0 & 21.6 \\
$B(E2:4^{+}_{1}\rightarrow2^{+}_{1})$ & 5.6 & 5.7 & 2.5 & 2.0 \\
$Q(2^{+}_{1})$ & +5.0 & +4.8 & +4.4 & +4.1 \\
$Q(2^{+}_{2})$ & -11.4 & -9.8 & -9.3 & -8.6 \\
$Q(4^{+}_{1})$ & +10.8 & +9.1 & +8.3 & +7.5 
\end{tabular}
\end{ruledtabular}
\label{48Ca_BE}
\end{table}%

\begin{figure}[tb]
\begin{center}
\includegraphics[width=\columnwidth]{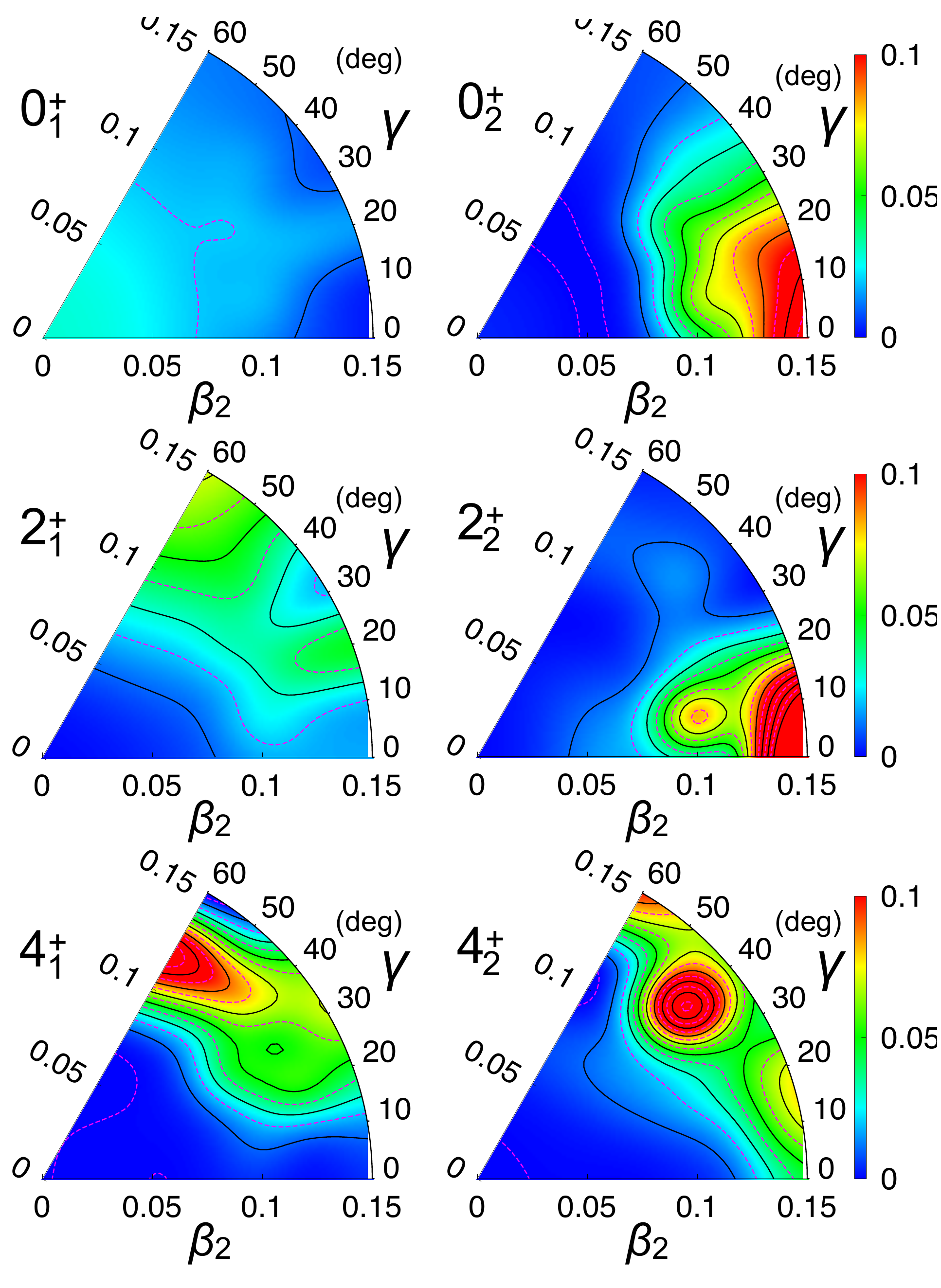}
\end{center}
\caption{(color online) Collective wave functions, $|F^\Gamma_\sigma (q)|^{2}$ (eq.~\eqref{coll_wf}) in the $q=(\beta_{2},\gamma)$ plane for the lowest $0^{+}$, $2^{+}$ and $4^{+}$ states in $^{48}$Ca computed with the PGCM$_{2}$ approximation. These functions are normalized to $\sum_{q}|F^\gamma_\sigma (q)|^{2}=1$. }
\label{Fig3bis}
\end{figure}
We now analyze two complementary aspects of the nuclear wave functions obtained by the PGCM framework, namely, the collective wave 
functions (c.w.f.) and the occupation numbers of the spherical single-particle orbits. The c.w.f.\@ are useful quantities to reveal the role of
the different collective degrees of freedom explored with constrained calculations and they are routinely computed 
in MREDF methods based on Skyrme, Gogny, and covariant functionals.
These functions give the most relevant 
contributions of the collective coordinates in each individual nuclear state~\cite{RS80a,Bender03a,Robledo18a} and are defined as
\begin{equation}
\left| F^{\Gamma}_{\sigma}(q)\right|^{2}= \bigg| \sum_{K\lambda} G^{\Gamma}_{\sigma;\lambda} u^{\Gamma}_{\lambda;qK} \bigg|^{2} .
\label{coll_wf}
\end{equation}
In Figure \ref{Fig3bis}, we plot the c.w.f.\@ for $0^{+}_{1,2}$, $2^{+}_{1,2}$ and $4^{+}_{1,2}$ states in the $(\beta_{2},\gamma)$ plane
computed with the PGCM$_{2}$ method. We first note that the ground state is flat in almost the whole range of deformations 
although the maximum contribution is found around the spherical point, as we could expect for a doubly-magic nucleus. 
This behavior is consistent with the degeneracy of the projected TES (Fig.~\ref{Fig1}(d)). The $2^{+}_{1}$ c.w.f.\@ is also rather smooth 
and we find a considerable mixing of oblate and $\gamma=20^{\circ}$ deformed states at the edges of the available $\beta_{2}$ 
deformations. The rest of the c.w.f.\@ represented in Fig.~\ref{Fig3bis} are also localized at (or nearby) the border of the $\beta_{2}$ 
coordinate. Moreover, they show a rather sharp behavior which means that only very 
few intrinsic states actually contribute to the building of the 
nuclear states. We recall that the limits in $\beta_{2}$-values are determined by the maximum value that gives a convergent PNVAP solution.
Finally, prolate deformed c.w.f.\@ with a similar 
structure are obtained for the $0^{+}_{2}$, $2^{+}_{2}$ and $4^{+}_{3}$ (not shown) states. However, the $4^{+}_{1}$, $4^{+}_{2}$ 
excited states are mainly located at specific $\gamma=60^{\circ}, 40^{\circ}$ values, respectively, and, to a lesser extent, at other $
\beta_{2}>0.1$ deformations.

It is important to note that prolate or oblate configurations of the $2^{+}$ and $4^{+}$ states obtained here are consistent with their values
for the spectroscopic quadrupole moments given in Table~\ref{48Ca_BE}. For example, the $2^{+}_{2}$ state is predominantly prolate
deformed and the $Q(2^{+}_{2})$-value is negative, and so on.  

\begin{figure}[tb]
\begin{center}
\includegraphics[width=\columnwidth]{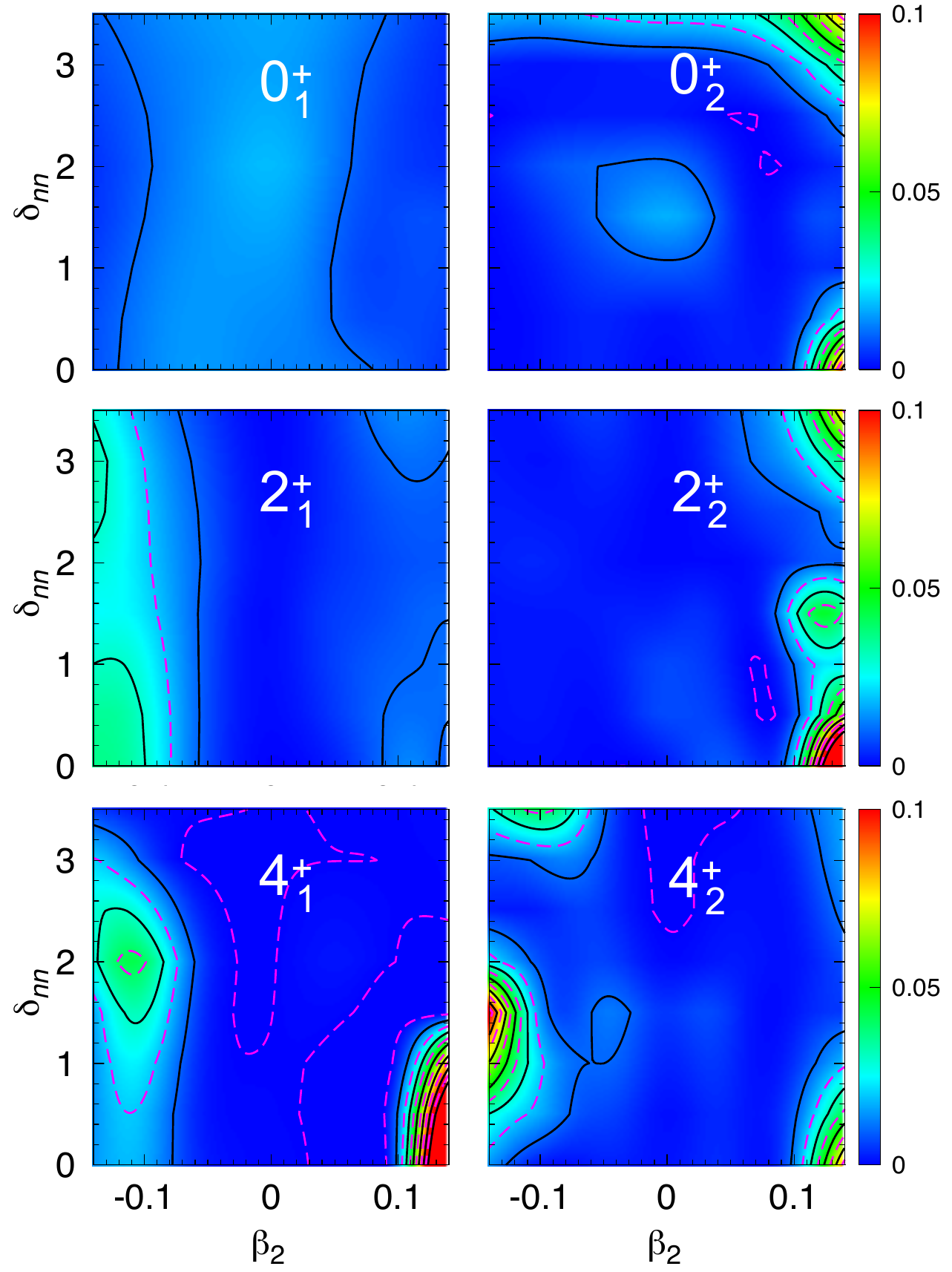}
\end{center}
\caption{(color online) Collective wave functions, $|F^\Gamma_\sigma (q)|^{2}$ (eq.~\eqref{coll_wf}) in the $q=(\beta_{2},\delta_{nn})$ 
plane for the lowest $0^{+}$, $2^{+}$ and $4^{+}$ states in $^{48}$Ca computed within the PGCM$_{3}$ approximation. 
These functions are normalized to $\sum_{q}|F^\Gamma_\sigma (q)|^{2}=1$. }
\label{Fig3tris}
\end{figure}

Similar conclusions can be extracted from the c.w.f.\@ given by the PGCM$_{3}$ method. In Fig.~\ref{Fig3tris} we show such quantities in 
the $(\beta_{2},\delta_{nn})$ plane with only axial deformed intrinsic states to simplify the discussion. The ground state is 
almost flat around $\beta_{2}\in\left[-0.1,0.1\right]$  with a mild dependence on $\delta_{nn}$. Again, the $0^{+}_{2}$ and $2^{+}_{2}$
c.w.f.\@ are functions located at the prolate edge with two sharp peaks, one at the largest and the other at the smallest pairing values. This 
topology is also seen in the $2^{+}_{1}$ c.w.f.\@ that is found at the oblate edge of the plot, consistently with the 
$2^{+}_{1}$ state obtained by the PGCM$_{2}$ approach discussed above. In fact, the only significant difference between the c.w.f.\@ 
calculated with the two approaches is found in the $4^{+}_{1}$ state. Here, both oblate states and a sharp region around prolate ($\beta_{2}=0.14$) and small-pairing values are present in this c.w.f.\@ while in the $(\beta_{2},\gamma)$ plane the oblate configuration is the predominant 
one.

Interestingly, these c.w.f.\@ (Figs.~\ref{Fig3bis}-~\ref{Fig3tris}) largely differ from those usually
obtained by EDF calculations where, in general, the distributions found are smoother and better behaved at the boundaries \cite{Bender03a,Niksic11a,Egido16a,Robledo18a}. 
One possible explanation is that in EDF calculations we have access to many more single-particle orbits that can evolve with the 
collective degrees of freedom following a Nilsson-like behavior. Such a picture allows for extending the exploration of these collective 
coordinates up to much larger values with the c.w.f.\@ decaying progressively to zero.  

More information can be learnt about the content of the correlated wave functions by computing their occupation numbers, i.e. the number of particles in a given $m_t$-orbit $\check{a} \equiv (\breve{a},m_t)$, 
\begin{equation}
n^{\Gamma\sigma}_{\check{a}} = \langle \Psi^{\Gamma}_{\sigma} \vert \hat{n}_{\check{a}} \vert \Psi^{\Gamma}_{\sigma} \rangle ,
\end{equation}
where
\begin{equation}
\hat{n}_{\check{a}} = \sum_{b} c^{\dagger}_{b} c_{b} \, \delta_{\breve{b} \breve{a}} \, \delta_{m_{t_b} m_t},
\end{equation}
and comparing the values to the decomposition of the exact eigenstates as obtained by the ISM calculations. A similar analysis was made in Ref.~\cite{Rodriguez16a} in the context of nuclear EDF methods. The results are shown in Table~\ref{Table_48OCC} for the nuclear states discussed above. 
The differences among the various PGCM schemes are small but the best agreement with the exact results is consistently given by the 
PGCM$_{3}$ approach. 
In fact, this approach is very close to the ISM results and reproduces very well the particle-hole structure of the 
ground and excited states. In particular, the ground state shows almost a pure $f_{7/2}$ closed-shell configuration, the $0^{+}_{2}$ and $2^{+}_{2}$
states are 2p-2h excitations in the $f_{7/2}$-$p_{3/2}$ orbits, the $2^{+}_{1}$ and $4^{+}_{1}$ states display 1p-1h configurations also 
in these orbits, and, finally, the $4^{+}_{2}$ state exhibits a 1p-1h configuration in the $f_{7/2}$-$f_{5/2}$ orbits. 
Therefore, the excitations 
found in this nucleus have, within our particular choice of basis, a well-defined single-particle nature. In addition, it is interesting to note that those kind of particle-hole configurations can be grasped by exploring \textit{a priori} collective degrees of freedom such as quadrupole deformations and pairing correlations. 
Actually, this observation together with the shapes of the collective wave functions displayed above seem to indicate that, in this small 
valence space, these variables are just useful tools to generate highly correlated intrinsic states and the meaning of "collectivity" is blurred. 
 
\begin{table}
\caption{Occupation numbers computed for the two lowest $0^{+}$, $2^{+}$ and $4^{+}$ states in $^{48}$Ca using the ISM method and 
the different PGCM schemes.}
\begin{ruledtabular}
\begin{tabular}{cccccc}
$J^{\pi\sigma}$ & orbit & PGCM$_{1}$ & PGCM$_{2}$ & PGCM$_{3}$ & ISM \\
\hline
$0^{+}_{1}$ & $f_{7/2}$ & 7.84 & 7.81 & 7.81 & 7.80 \\
& $p_{3/2}$ & 0.10 & 0.08 & 0.08 & 0.07 \\
& $f_{5/2}$ & 0.04 & 0.09 & 0.09 & 0.11 \\
& $p_{1/2}$ & 0.02 & 0.02 & 0.02 & 0.02 \\
\hline
$0^{+}_{2}$ & $f_{7/2}$ & 5.89 & 5.84 & 5.84 & 5.82 \\
& $p_{3/2}$ & 1.44 & 1.64 & 1.75 & 1.80 \\
& $f_{5/2}$ & 0.15 & 0.18 & 0.15 & 0.17 \\
& $p_{1/2}$ & 0.52 & 0.34 & 0.26 & 0.21 \\
\hline
$2^{+}_{1}$ & $f_{7/2}$ & 6.80 & 6.77 & 6.77 & 6.77 \\
& $p_{3/2}$ & 1.07 & 1.07 & 1.05 & 1.08 \\
& $f_{5/2}$ & 0.09 & 0.12 & 0.14 & 0.11 \\
& $p_{1/2}$ & 0.04 & 0.04 & 0.04 & 0.04 \\
\hline
$2^{+}_{2}$ & $f_{7/2}$ & 5.86 & 5.85 & 5.87 & 5.85 \\
& $p_{3/2}$ & 1.44 & 1.63 & 1.71 & 1.78 \\
& $f_{5/2}$ & 0.15 & 0.14 & 0.12 & 0.13 \\
& $p_{1/2}$ & 0.55 & 0.38 & 0.30 & 0.24 \\
\hline
$4^{+}_{1}$ & $f_{7/2}$ & 6.66 & 6.59 & 6.83 & 6.83 \\
& $p_{3/2}$ & 1.15 & 1.12 & 1.06 & 1.03 \\
& $f_{5/2}$ & 0.13 & 0.23 & 0.08 & 0.11 \\
& $p_{1/2}$ & 0.06 & 0.06 & 0.03 & 0.03 \\
\hline
$4^{+}_{2}$ & $f_{7/2}$ & 6.63 & 6.60 & 6.79 & 6.82 \\
& $p_{3/2}$ & 0.43 & 0.37 & 0.17 & 0.09 \\
& $f_{5/2}$ & 0.11 & 0.15 & 0.10 & 0.11 \\
& $p_{1/2}$ & 0.83 & 0.88 & 0.94 & 0.98
\end{tabular}
\end{ruledtabular}
\label{Table_48OCC}
\end{table}%

\begin{figure}[tb]
\begin{center}
\includegraphics[width=0.8\columnwidth]{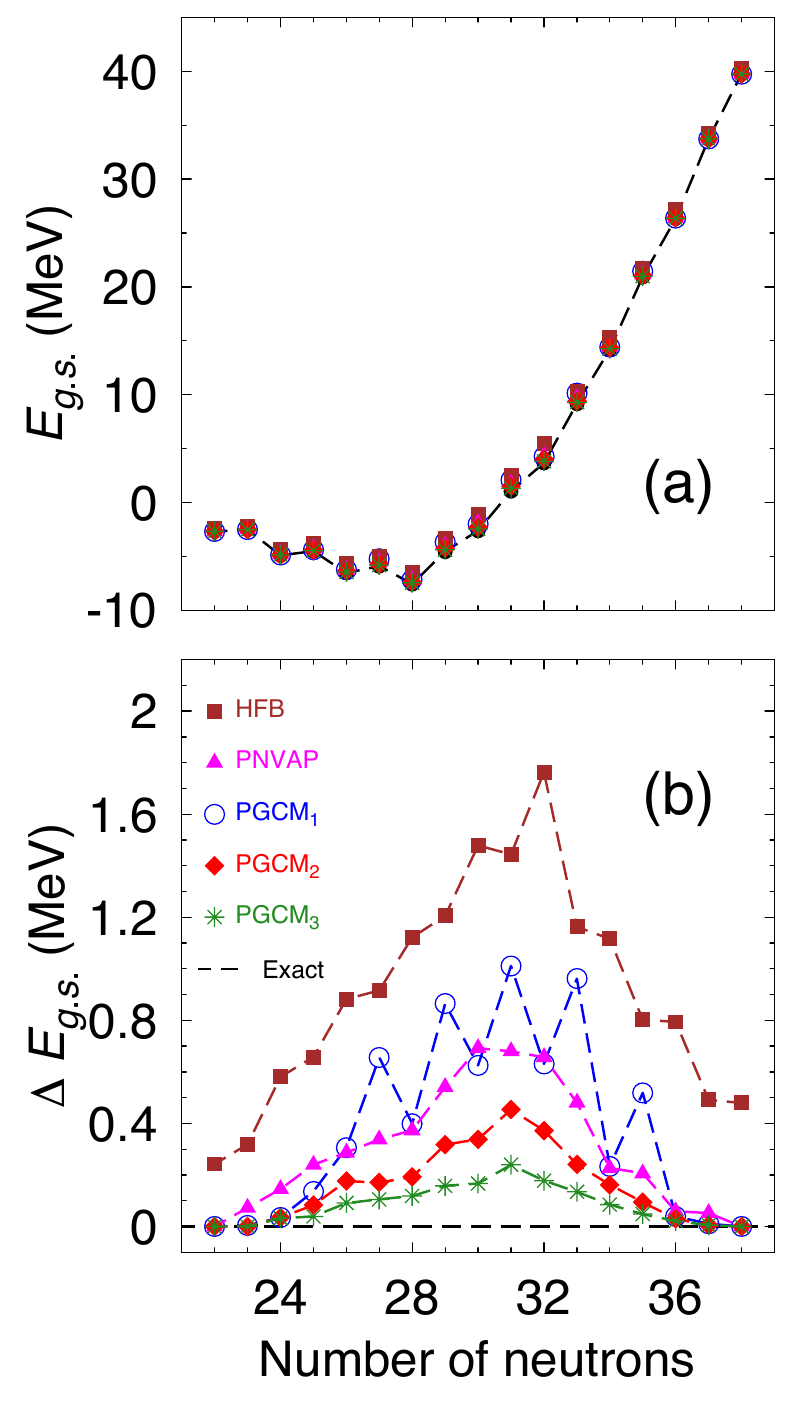}
\end{center}
\caption{(color online) (a) Ground state energy and (b) energy difference between the approximate and exact ground state energies 
computed for the calcium isotopic chain ($^{42-60}$Ca) with different variational approaches (HFB (brown filled boxes), PNVAP 
(magenta filled triangles), PGCM$_{1}$ (blue circles), PGCM$_{2}$ (red filled diamonds) and PGCM$_{3}$ (green asterisks)). Exact 
results (ISM) are represented by a black dashed line.}
\label{Fig4}
\end{figure}
\subsection{Calcium isotopic chain} 

We now turn our attention towards the results obtained along the whole calcium isotopic chain in the $pf$-shell. In Fig.~\ref{Fig4}(a) we represent the ground-state 
energies for even-even and odd-even calcium nuclei computed exactly and with the different variational schemes. We see that the results are
almost indistinguishable at the scale that reflects the absolute ground-state energies. To better visualize the 
accuracy of the different variational methods, we plot the difference between the approximate and exact energies, 
$\Delta E_{g.s.}=E^{approx}_{g.s.}-E^{ISM}_{g.s.}$, in Fig.~\ref{Fig4}(b). Here, we observe a better agreement with the exact result at 
the edges of the valence space than in the middle. As we will discuss below, the dimensions of the Hamiltonian matrices are the largest in the 
mid-shell and there will be configurations that cannot be 
captured only exploring the collective degrees of freedom studied here. The worst approximation is given by the HFB method, as
expected, where a slight odd-even staggering is found. The smallest (largest) difference is $0.24$ MeV ($1.76$ MeV) obtained in 
the nucleus $^{42}$Ca ($^{52}$Ca). The staggering is magnified in the results provided by the PGCM$_{1}$ method that
uses precisely the HFB intrinsic wave functions to define the subspace within which the GCM states are constructed. 
In this case, the even nuclei are better described than their odd
neighbors except for the isotopes at the edges ($^{42-45,56-58}$Ca) where the PGCM$_{1}$ approximation gives ground state energies
very close to the exact solution (for $^{42,43,57,58}$Ca they are on top of the ISM results). Using the PGCM$_{1}$ method, the largest difference is 
reduced to $1.01$ MeV found in the isotope $^{51}$Ca. Smaller and smoother energy differences are obtained by 
PNVAP approaches (PNVAP itself and PGCM$_{2,3}$).
In particular, the superior treatment of pairing correlations in PNVAP approaches allows them to describe even- and odd-mass isotopes on an equal footing, i.e.\@ there is no odd-even staggering in those cases.
By constrast, HFB is a poor approximation whenever pairing correlations are weak, e.g.\@ odd number parity states due to the blocking effect.
In the best case, namely, the PGCM$_{3}$ scheme, we obtain a largest 
difference of $0.24$ MeV (at $^{51}$Ca) proving the ability of these variational methods to approach closely the exact ground states.

\begin{figure*}[t]
\begin{center}
\includegraphics[width=0.8\textwidth]{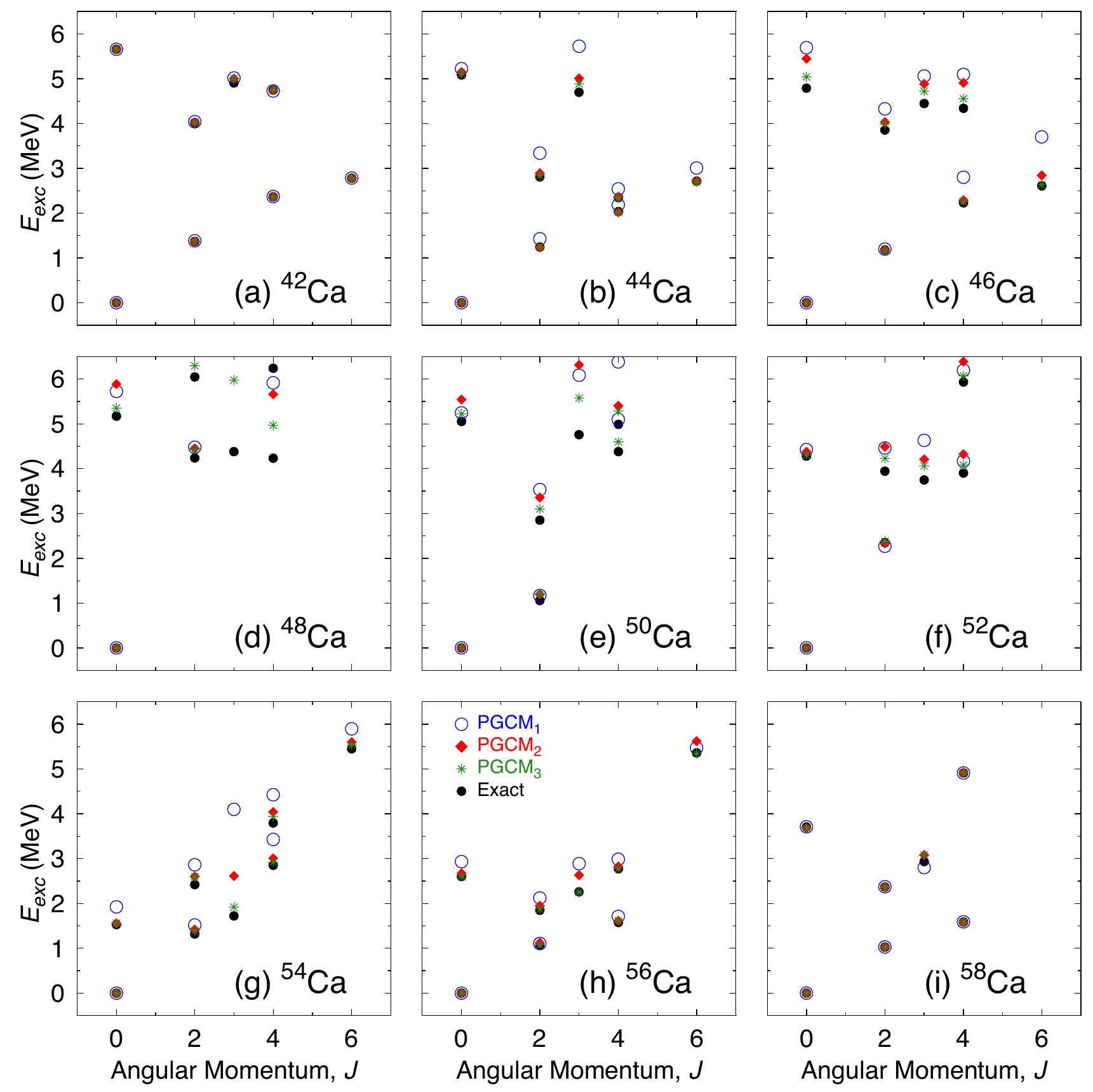}
\end{center}
\caption{(color online) Excitation energies as a function of the angular momentum 
calculated with ISM (black bullets), PGCM$_{1}$ (blue circles), PGCM$_{2}$ (red filled diamonds)
and PGCM$_{3}$ (green asterisks) for even calcium isotopes.}
\label{Fig5}
\end{figure*}
\begin{figure}[tb]
\begin{center}
\includegraphics[width=\columnwidth]{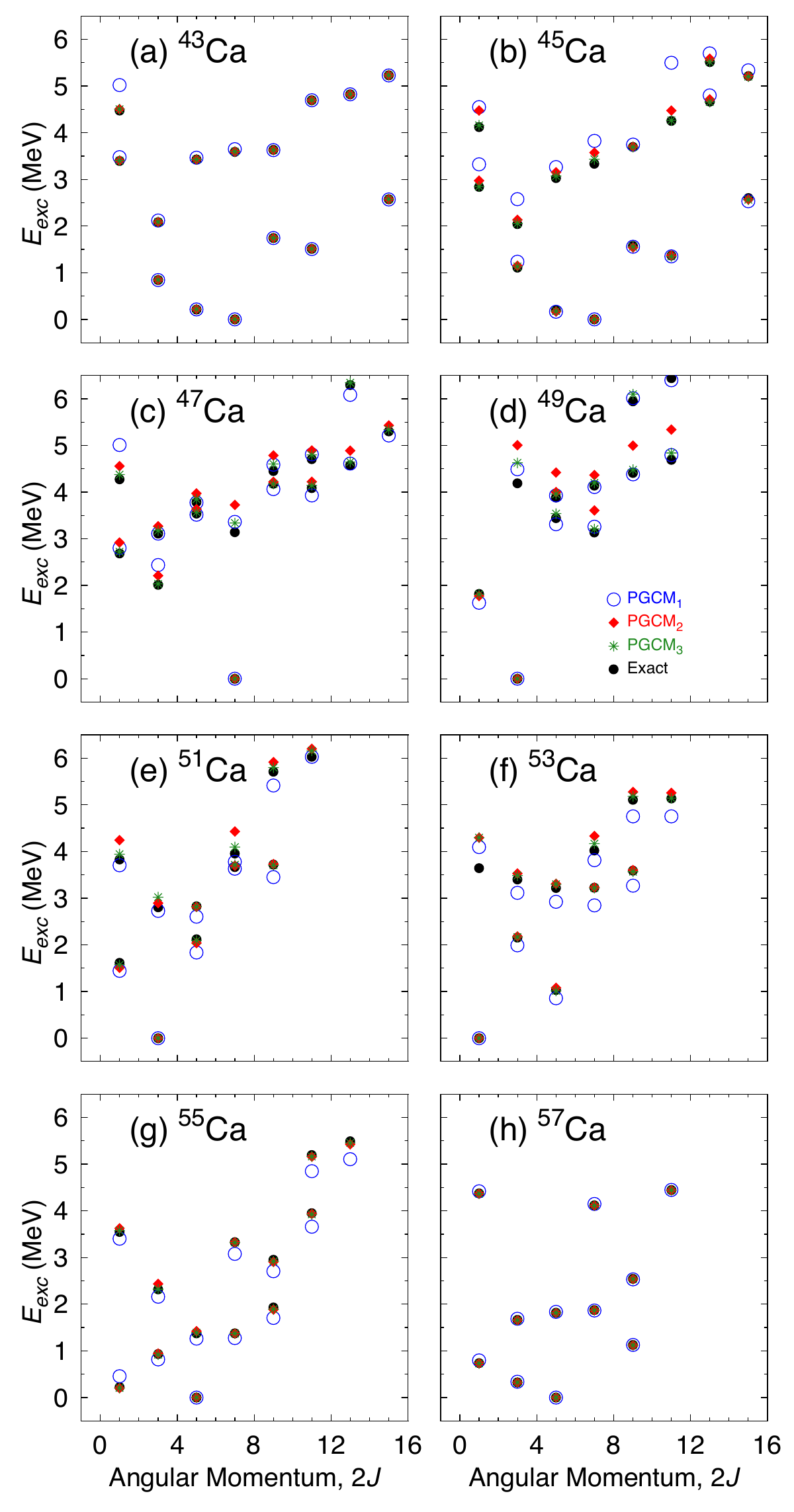}
\end{center}
\caption{(color online) Same as Fig.~\ref{Fig5} but for odd calcium isotopes.} 
\label{Fig6}
\end{figure}
We can also analyze the capability of these methods to reproduce the low-lying excitation spectra. In that order, the excitation energies of the lowest states computed with ISM and PGCM methods are displayed as a function of the angular momentum in Fig.~\ref{Fig5} and Fig.~\ref{Fig6} for even and odd calcium isotopes, respectively. The overall agreement with the exact results is again very good
and the quantitative behavior of the excitation energies is well reproduced. In fact, the ISM solutions are obtained exactly at the edges of 
the valence space even with PGCM$_{1,2}$ approaches. The quantitative agreement deteriorates in the mid-shell but it is improved by the 
inclusion of the pairing fluctuations within the PGCM$_{3}$ method. We also notice in that case a better description of the exact spectra in odd-mass nuclei.
The effect of pairing on the final spectra can be studied by comparing the PGCM approximations based on the HFB solutions with those 
based on the PNVAP ones. For even-mass nuclei, the PGCM$_{1}$ approach produces systematically larger excitation energies 
in even nuclei that are subsequently reduced in the PGCM$_{2,3}$ methods. For odd-mass nuclei, the PGCM$_{2}$ approach gives 
systematically larger excitation energies that are corrected once the pairing fluctuations are included. Consequently, the PGCM$_{3}$ 
calculations provide almost the exact spectra in the odd calcium isotopes. 

\begin{figure}[tb]
\begin{center}
\includegraphics[width=\columnwidth]{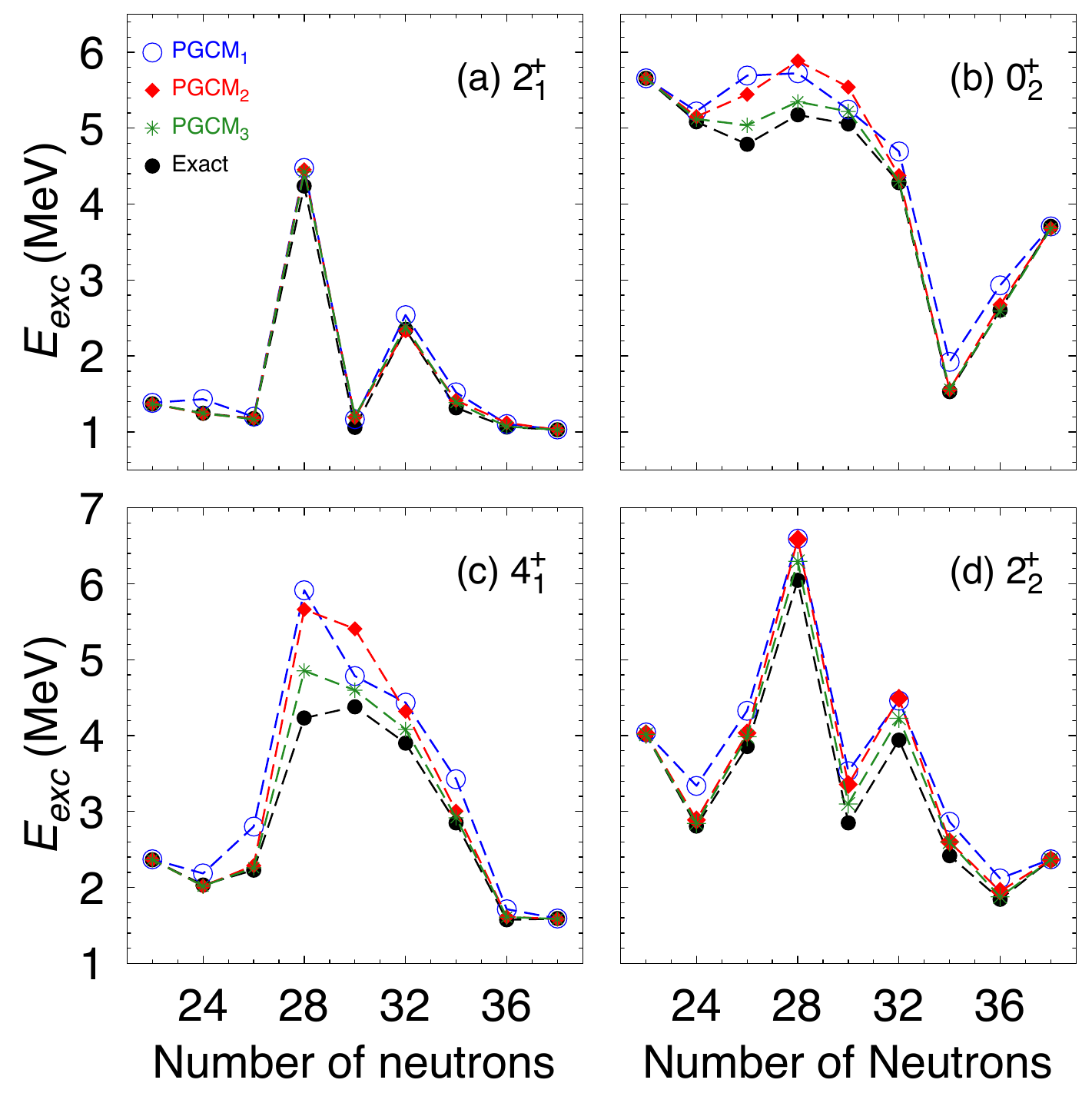}
\end{center}
\caption{(color online) Excitation energies for even calcium isotopes in the $pf$-shell computed exactly and with
PGCM$_{1,2,3}$ variational approximations.} 
\label{Fig7}
\end{figure}

For even-mass isotopes, we display in Fig.~\ref{Fig7} the systematics of the lowest excitations with an even value of $J$. Again, we observe a very good reproduction of the exact results by the PGCM methods. In particular, these approximations are able to reproduce the main trends, e.g., the peaks at $N=28$ and $32$ in the $2^{+}_{1,2}$ 
states associated to shell closures, the sudden decrease of the $0^{+}_{2}$ energy at $N=34$, or the bell-shaped $4^{+}_{1}$ excitation 
energies. However, the PGCM$_{1,2}$ approximations have problems in describing the $0^{+}_{2}$ and $4^{+}_{1}$
excitation energies in the vicinity of $^{48}$Ca. This discrepancy is partially corrected by the addition of the pairing fluctuations 
(PGCM$_{3}$) although there is some room left to improvements, 
which could be corrected through the inclusion of additional degrees of freedom such as the rotational frequency or the explicit multi-quasi-particle excitations. Nevertheless, the PGCM$_{3}$ approximation is almost on top of the exact results in most of the excitation 
energies shown in Fig.~\ref{Fig7}.

\begin{figure*}[t]
\begin{center}
\includegraphics[width=0.8\textwidth]{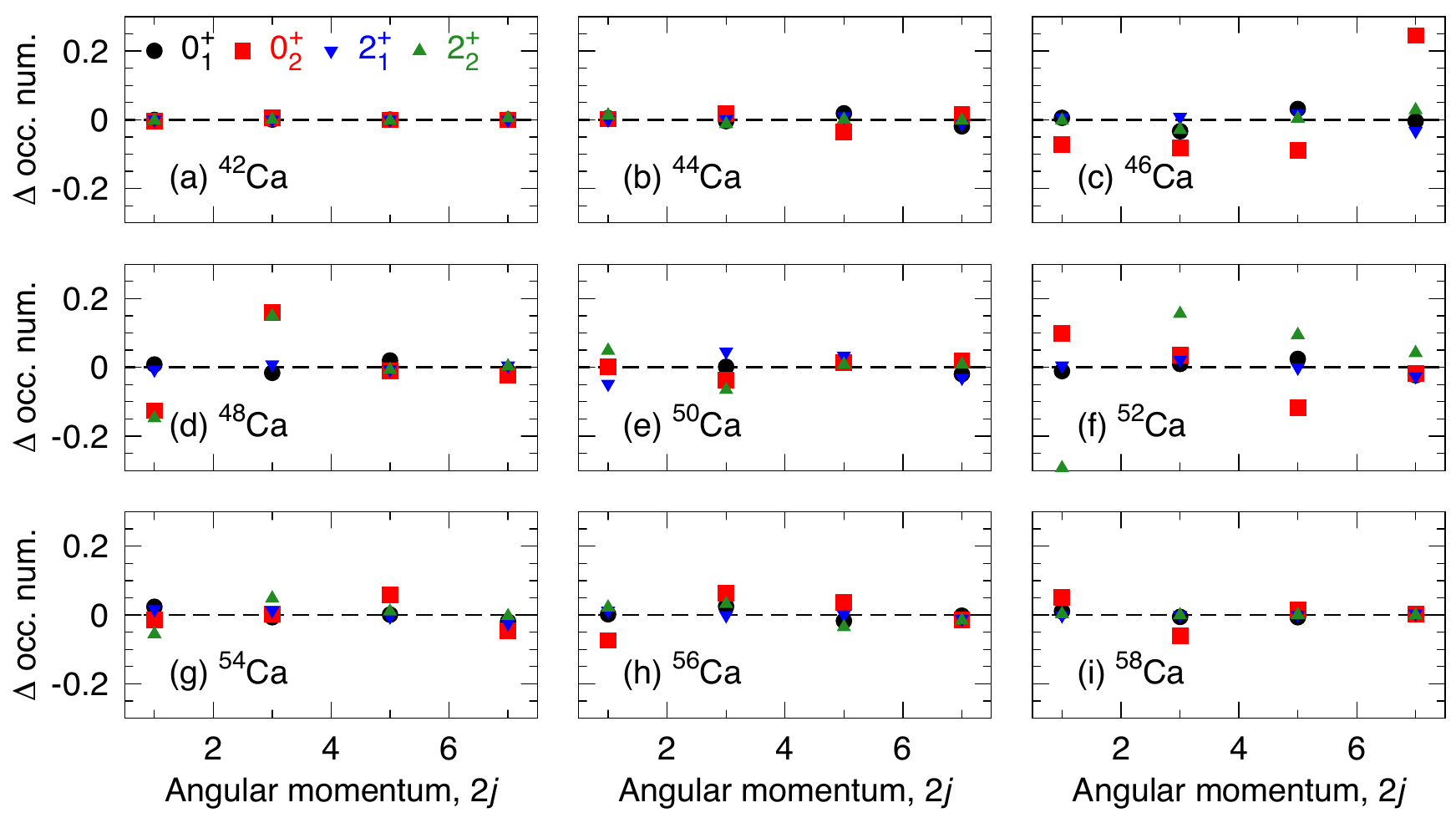}
\end{center}
\caption{(color online) Difference between the occupation numbers computed with the ISM and the PGCM$_{2}$ methods for all even 
calcium isotopes in the $pf$-shell. Ground (black bullets), $0^{+}_{2}$ (red filled boxes), $2^{+}_{1}$ (blue triangles down) and $2^{+}_{2}$ 
(green triangles up) are represented.} 
\label{Fig8}
\end{figure*}

The study of the occupation numbers performed above for $^{48}$Ca can be extended to all even 
calcium isotopes. For the sake of simplicity, we only represent in Fig.~\ref{Fig8} the difference between the ISM and the PGCM$_{2}$ 
methods for the two first $0^{+}$ and $2^{+}$ states. We observe that the difference for the ground and the $2^{+}_{1}$ 
states are almost negligible in the whole isotopic chain. 
Slightly larger differences are found for $0^{+}_{2}$ and $2^{+}_{2}$ in the mid-shell nuclei ($^{46-52}$Ca), consistently with the 
discrepancies shown above for ground and excited state energies. However, these differences are smaller than 0.2 particles in most of the 
cases. Therefore, these results prove the ability of the PGCM methods to describe the underlying single-particle structure of the 
nuclear wave functions by exploring coordinates that depict, \textit{a priori}, collective degrees of freedom.    

These results suggest that, for all the values of the angular momentum $J$ examined here, the lowest eigenstates of the full many-body Hilbert $J$-subspaces can be well approximated by the variational mixing of few selected correlated many-body states. As described in Sec.~\ref{Theoretical_Framework}, our approximated wave functions are built as optimal superpositions of the symmetry-projected quasi-particle states included in our trial set. This is equivalent to the diagonalization of the Hamiltonian in the subspace spanned by the set of projected states $\left\{ P^{J}_{MK}P^{N}P^{Z}P^{\pi}|\Phi(q)\rangle , \forall \, qK \right\}$. In practice, however, due to the linear redundancy among the projected states, the dimension of the subspace within which the Hamiltonian is effectively diagonalized is much smaller.
In Table~\ref{Table_dim}, we compare the dimensions of the exact diagonalization 
using Slater determinants coupled to $J$ with the dimensions of the GCM diagonalization using the natural basis states for even and odd calcium isotopes (the dimensions are symmetric with respect to 
particles and holes). 
For simplicity, we compare again only the PGCM$_{2}$ and ISM methods. The number of intrinsic states, $|
\Phi(\beta_{2},\gamma)\rangle$, is 50 in this case. Since $K$-mixing is also performed, the actual number of states mixed within the GCM 
framework for each angular momentum is $50\times(2J+1)$. We observe in Table~\ref{Table_dim} a large reduction of this number 
whenever the corresponding natural bases are extracted, i.e., a substantial redundancy given by the linear dependence of the projected 
states is obtained. It is important to point out that at the edges of the valence space ($^{42-58}$Ca, $^{43-57}$Ca) the dimensions of the 
natural bases are the same or almost the same as the bases built with Slater determinants coupled to $J$. 
\begin{table}[tb]
\caption{Dimensions of the Hamiltonian matrices for ISM (value on the left) and PGCM$_{2}$ (value on the right) in even (top) and odd (bottom) 
calcium isotopes in the $pf$-shell.}
\begin{ruledtabular}
\begin{tabular}{cccccc}
$J$ & 42 or 58 & 44 or 56 & 46 or 54 & 48 or 52 & 50 \\
\hline
0 & 4 $\vert$ 4 & 28 $\vert$  7$\phantom{0}$ & 137 $\vert$ 7$\phantom{0}$ & 347 $\vert$  9 & 468 $\vert$  9 \\
2 & 8 $\vert$  7 & 94 $\vert$  16 & 512 $\vert$  14 & 1390 $\vert$  20 & 1935 $\vert$  20 \\
4 & 6 $\vert$  6 & 99 $\vert$  21 & 615 $\vert$  20 & 1755 $\vert$  28 & 2468 $\vert$  13 \\
6 & 2 $\vert$  2 & 59 $\vert$  17 & 462 $\vert$  17 & 1426 $\vert$  26 &  2051 $\vert$  8$\phantom{0}$ \\[0.5mm]
\hline
\hline
$J$ & 43 or 57 & 45 or 55 & 47 or 53 & 49 or 51 & \\
\hline
1/2 & 12 $\vert$ 9$\phantom{0}$ & 107 $\vert$  10 & 415 $\vert$  12 & $\phantom{0}$790 $\vert$  12 &  \\
3/2 & 25 $\vert$ 18 & 198 $\vert$  21 & 764 $\vert$  26 & 1484 $\vert$  15 &  \\
5/2 & 28 $\vert$ 21 & 253 $\vert$  25 & 1005 $\vert$  37$\phantom{0}$ & 1965 $\vert$  30 & \\
7/2 & 27 $\vert$ 21 & 271 $\vert$  23 & 1121 $\vert$  24$\phantom{0}$ & 2215 $\vert$  32 & \\
9/2 & 23 $\vert$ 22 & 252 $\vert$  41 & 1091 $\vert$  45$\phantom{0}$ & 2214 $\vert$  38 & \\
11/2 & 16 $\vert$ 16 & 211 $\vert$  34 & 974 $\vert$  45 & 2017 $\vert$  35 & \\
13/2 & 8 $\vert$ 8 & 153 $\vert$  33 & 783 $\vert$  39 & 1669 $\vert$  28 & \\
15/2 & 5 $\vert$ 5 & 105 $\vert$  25 & 577 $\vert$  33 & 1284 $\vert$  21 & 
\end{tabular}
\end{ruledtabular}
\label{Table_dim}
\end{table}%
Therefore, in those cases the ISM and PGCM 
methods are necessarily equivalent and that explains the perfect agreement between exact diagonalizations and the variational approaches. 
However, for systems in the mid-shell, the number of many-body states in the natural bases are two 
orders of magnitude smaller than the shell-model bases and still the description is very good. Some pros and cons of this procedure can be 
mentioned. The main advantage of the PGCM methods are their scalability with the number of configurations/shells in the valence space. 
In this sense, these methods can be used in a broader context. We also see that these techniques 
are straightforwardly extendable by including additional degrees of freedom (pairing fluctuations, cranking terms, explicit quasi-particle 
excitations), even though the ground states are rather well approximated by including only the most relevant variables (quadrupole and 
pairing). However, the addition of more constraints is also costly from a computational point of view and the method can be eventually as 
complicated as the full diagonalization. We also observe a poorer description of the exact results in those systems where the number of 
Slater determinants is very much larger than the dimension of the corresponding natural basis (e.g., states in $^{46-54}$Ca). Concerning 
the latter point, it is obvious that only few (out of the total)  number of nuclear states for a given angular momentum can be obtained by the 
current implementations of the PGCM method. Nevertheless, the present study shows 
that these methods are able to reproduce very well the exact results by mixing a very reduced set of strongly correlated basis states. 

\section{Summary and outlook}
\label{Summary}

In this article, we have studied the performance of several PGCM approximation schemes in reproducing the exact eigenstates of the shell model Hamiltonian KB3G in the $pf$-shell valence space. The PGCM  is a versatile variational method that includes beyond-mean-field correlations through the configuration mixing of a set symmetry-projected intrinsic states. In this work, the intrinsic states considered are Bogolyubov quasi-particle states that explore the quadrupole and the pairing degrees of freedom, and they have been generated by solving a series of constrained HFB or PNVAP equations.

We used the even and odd calcium isotopes in the $pf$-shell valence space to benchmark the different PGCM implementations. In general, we obtained a very good qualitative and quantitative 
description of ground- and excited-state exact energies. 
Furthermore, the use of PNVAP intrinsic states and the simultaneous exploration of 
quadrupole and $nn$-pairing degrees of freedom improved significantly the agreement with the shell model results, especially in the odd systems. 

We have also analyzed the PGCM correlated wave functions in terms of the collective coordinates and their occupation numbers. For the nucleus $^{48}$Ca, we have found a very flat ground-state collective wave function and 
sharply peaked excited-state collective wave functions. These distributions seem to indicate that the meaning of the collective degrees of
freedom in these isotopes and valence spaces is not as clear as in well-deformed nuclei and no-core configuration spaces.
Recently, a similar conclusion has been drawn after analyzing the $(\beta_{2},\gamma)$ values and their variances extracted from the Kumar 
invariants~\cite{Poves19a} in spherical isotopes. The non-collective nature of the excitations in this isotopic chain could nonetheless be reproduced 
by the present PGCM implementations as the analysis of the occupation numbers pointed out.   

Altogether, our PGCM calculations have demonstrated that a very good approximation to the lowest eigenstates of the shell-model Hamiltonian can be obtained through the diagonalization of the aforementioned Hamiltonian within the subspace spanned by a small set of selected correlated wave functions instead of very large $J$-coupled Slater determinant basis. This is in line with the conclusions drawn in Refs.~\cite{Ripoche17a,Ripoche18a} in the context of a truncated configuration-interaction method.
There were, however, some discrepancies for some states belonging to many-body Hilbert spaces with large dimensions (i.e.\@ for mid-shell isotopes). In those cases, the exploration of additional degrees of freedom, such as the rotational frequency or the explicit multi-quasi-particle excitations, in the generation of the intrinsic states would be needed. This is a work in progress. 

It would certainly be interesting to extend the present study by testing the quality of our variational approximations in a larger valence space that includes more than a single harmonic oscillator shell. Finally, the role of other pairing channels, in particular, the isoscalar and isovector proton-neutron pairing degrees of freedom will be studied in the near future.       

\section*{Acknowledgements}

We thank L. M. Robledo for proofreading the manuscript.  We also thank him, A. Poves, and 
 F. Nowacki for useful discussions. 
The work of TRR was supported by the Spanish MINECO under Programa Ram\'on y Cajal 11420, FIS-2014-53434-P and
by the Spanish MICINN under PGC2018-094583-B-I00, and, in part, by the ExtreMe Matter Institute EMMI
at the GSI-Darmstadt, Germany. TRR gratefully thanks the support from GSI-Darmstadt 
and TU-Darmstadt computing facilities.


\bibliography{biblio}

\begin{thebibliography}{57}%
\makeatletter
\providecommand \@ifxundefined [1]{%
 \@ifx{#1\undefined}
}%
\providecommand \@ifnum [1]{%
 \ifnum #1\expandafter \@firstoftwo
 \else \expandafter \@secondoftwo
 \fi
}%
\providecommand \@ifx [1]{%
 \ifx #1\expandafter \@firstoftwo
 \else \expandafter \@secondoftwo
 \fi
}%
\providecommand \natexlab [1]{#1}%
\providecommand \enquote  [1]{``#1''}%
\providecommand \bibnamefont  [1]{#1}%
\providecommand \bibfnamefont [1]{#1}%
\providecommand \citenamefont [1]{#1}%
\providecommand \href@noop [0]{\@secondoftwo}%
\providecommand \href [0]{\begingroup \@sanitize@url \@href}%
\providecommand \@href[1]{\@@startlink{#1}\@@href}%
\providecommand \@@href[1]{\endgroup#1\@@endlink}%
\providecommand \@sanitize@url [0]{\catcode `\\12\catcode `\$12\catcode
  `\&12\catcode `\#12\catcode `\^12\catcode `\_12\catcode `\%12\relax}%
\providecommand \@@startlink[1]{}%
\providecommand \@@endlink[0]{}%
\providecommand \url  [0]{\begingroup\@sanitize@url \@url }%
\providecommand \@url [1]{\endgroup\@href {#1}{\urlprefix }}%
\providecommand \urlprefix  [0]{URL }%
\providecommand \Eprint [0]{\href }%
\providecommand \doibase [0]{http://dx.doi.org/}%
\providecommand \selectlanguage [0]{\@gobble}%
\providecommand \bibinfo  [0]{\@secondoftwo}%
\providecommand \bibfield  [0]{\@secondoftwo}%
\providecommand \translation [1]{[#1]}%
\providecommand \BibitemOpen [0]{}%
\providecommand \bibitemStop [0]{}%
\providecommand \bibitemNoStop [0]{.\EOS\space}%
\providecommand \EOS [0]{\spacefactor3000\relax}%
\providecommand \BibitemShut  [1]{\csname bibitem#1\endcsname}%
\let\auto@bib@innerbib\@empty
\bibitem [{\citenamefont {Epelbaum}\ \emph {et~al.}(2009)\citenamefont
  {Epelbaum}, \citenamefont {Hammer},\ and\ \citenamefont
  {Mei\ss{}ner}}]{Epelbaum09a}%
  \BibitemOpen
  \bibfield  {author} {\bibinfo {author} {\bibfnamefont {E.}~\bibnamefont
  {Epelbaum}}, \bibinfo {author} {\bibfnamefont {H.-W.}\ \bibnamefont
  {Hammer}}, \ and\ \bibinfo {author} {\bibfnamefont {U.-G.}\ \bibnamefont
  {Mei\ss{}ner}},\ }\href {\doibase 10.1103/RevModPhys.81.1773} {\bibfield
  {journal} {\bibinfo  {journal} {Rev. Mod. Phys.}\ }\textbf {\bibinfo {volume}
  {81}},\ \bibinfo {pages} {1773} (\bibinfo {year} {2009})}\BibitemShut
  {NoStop}%
\bibitem [{\citenamefont {Hammer}\ \emph {et~al.}(2019)\citenamefont {Hammer},
  \citenamefont {König},\ and\ \citenamefont {van Kolck}}]{Hammer19a}%
  \BibitemOpen
  \bibfield  {author} {\bibinfo {author} {\bibfnamefont {H.~W.}\ \bibnamefont
  {Hammer}}, \bibinfo {author} {\bibfnamefont {S.}~\bibnamefont {König}}, \
  and\ \bibinfo {author} {\bibfnamefont {U.}~\bibnamefont {van Kolck}},\
  }\href@noop {} {\  (\bibinfo {year} {2019})},\ \Eprint
  {http://arxiv.org/abs/1906.12122} {arXiv:1906.12122 [nucl-th]} \BibitemShut
  {NoStop}%
\bibitem [{\citenamefont {Som\`a}\ \emph {et~al.}(2013)\citenamefont {Som\`a},
  \citenamefont {Barbieri},\ and\ \citenamefont {Duguet}}]{Soma13a}%
  \BibitemOpen
  \bibfield  {author} {\bibinfo {author} {\bibfnamefont {V.}~\bibnamefont
  {Som\`a}}, \bibinfo {author} {\bibfnamefont {C.}~\bibnamefont {Barbieri}}, \
  and\ \bibinfo {author} {\bibfnamefont {T.}~\bibnamefont {Duguet}},\ }\href
  {\doibase 10.1103/PhysRevC.87.011303} {\bibfield  {journal} {\bibinfo
  {journal} {Phys. Rev. C}\ }\textbf {\bibinfo {volume} {87}},\ \bibinfo
  {pages} {011303} (\bibinfo {year} {2013})}\BibitemShut {NoStop}%
\bibitem [{\citenamefont {Hagen}\ \emph {et~al.}(2014)\citenamefont {Hagen},
  \citenamefont {Papenbrock}, \citenamefont {Hjorth-Jensen},\ and\
  \citenamefont {Dean}}]{Hagen14a}%
  \BibitemOpen
  \bibfield  {author} {\bibinfo {author} {\bibfnamefont {G.}~\bibnamefont
  {Hagen}}, \bibinfo {author} {\bibfnamefont {T.}~\bibnamefont {Papenbrock}},
  \bibinfo {author} {\bibfnamefont {M.}~\bibnamefont {Hjorth-Jensen}}, \ and\
  \bibinfo {author} {\bibfnamefont {D.~J.}\ \bibnamefont {Dean}},\ }\href
  {\doibase 10.1088/0034-4885/77/9/096302} {\bibfield  {journal} {\bibinfo
  {journal} {Reports on Progress in Physics}\ }\textbf {\bibinfo {volume}
  {77}},\ \bibinfo {pages} {096302} (\bibinfo {year} {2014})}\BibitemShut
  {NoStop}%
\bibitem [{\citenamefont {Hergert}\ \emph {et~al.}(2016)\citenamefont
  {Hergert}, \citenamefont {Bogner}, \citenamefont {Morris}, \citenamefont
  {Schwenk},\ and\ \citenamefont {Tsukiyama}}]{Hergert16a}%
  \BibitemOpen
  \bibfield  {author} {\bibinfo {author} {\bibfnamefont {H.}~\bibnamefont
  {Hergert}}, \bibinfo {author} {\bibfnamefont {S.}~\bibnamefont {Bogner}},
  \bibinfo {author} {\bibfnamefont {T.}~\bibnamefont {Morris}}, \bibinfo
  {author} {\bibfnamefont {A.}~\bibnamefont {Schwenk}}, \ and\ \bibinfo
  {author} {\bibfnamefont {K.}~\bibnamefont {Tsukiyama}},\ }\href {\doibase
  https://doi.org/10.1016/j.physrep.2015.12.007} {\bibfield  {journal}
  {\bibinfo  {journal} {Physics Reports}\ }\textbf {\bibinfo {volume} {621}},\
  \bibinfo {pages} {165 } (\bibinfo {year} {2016})},\ \bibinfo {note} {memorial
  Volume in Honor of Gerald E. Brown}\BibitemShut {NoStop}%
\bibitem [{\citenamefont {Tichai}\ \emph {et~al.}(2018)\citenamefont {Tichai},
  \citenamefont {Arthuis}, \citenamefont {Duguet}, \citenamefont {Hergert},
  \citenamefont {Som\`a},\ and\ \citenamefont {Roth}}]{Tichai18a}%
  \BibitemOpen
  \bibfield  {author} {\bibinfo {author} {\bibfnamefont {A.}~\bibnamefont
  {Tichai}}, \bibinfo {author} {\bibfnamefont {P.}~\bibnamefont {Arthuis}},
  \bibinfo {author} {\bibfnamefont {T.}~\bibnamefont {Duguet}}, \bibinfo
  {author} {\bibfnamefont {H.}~\bibnamefont {Hergert}}, \bibinfo {author}
  {\bibfnamefont {V.}~\bibnamefont {Som\`a}}, \ and\ \bibinfo {author}
  {\bibfnamefont {R.}~\bibnamefont {Roth}},\ }\href {\doibase
  https://doi.org/10.1016/j.physletb.2018.09.044} {\bibfield  {journal}
  {\bibinfo  {journal} {Physics Letters B}\ }\textbf {\bibinfo {volume}
  {786}},\ \bibinfo {pages} {195 } (\bibinfo {year} {2018})}\BibitemShut
  {NoStop}%
\bibitem [{\citenamefont {Morris}\ \emph {et~al.}(2018)\citenamefont {Morris},
  \citenamefont {Simonis}, \citenamefont {Stroberg}, \citenamefont {Stumpf},
  \citenamefont {Hagen}, \citenamefont {Holt}, \citenamefont {Jansen},
  \citenamefont {Papenbrock}, \citenamefont {Roth},\ and\ \citenamefont
  {Schwenk}}]{Morris18a}%
  \BibitemOpen
  \bibfield  {author} {\bibinfo {author} {\bibfnamefont {T.~D.}\ \bibnamefont
  {Morris}}, \bibinfo {author} {\bibfnamefont {J.}~\bibnamefont {Simonis}},
  \bibinfo {author} {\bibfnamefont {S.~R.}\ \bibnamefont {Stroberg}}, \bibinfo
  {author} {\bibfnamefont {C.}~\bibnamefont {Stumpf}}, \bibinfo {author}
  {\bibfnamefont {G.}~\bibnamefont {Hagen}}, \bibinfo {author} {\bibfnamefont
  {J.~D.}\ \bibnamefont {Holt}}, \bibinfo {author} {\bibfnamefont {G.~R.}\
  \bibnamefont {Jansen}}, \bibinfo {author} {\bibfnamefont {T.}~\bibnamefont
  {Papenbrock}}, \bibinfo {author} {\bibfnamefont {R.}~\bibnamefont {Roth}}, \
  and\ \bibinfo {author} {\bibfnamefont {A.}~\bibnamefont {Schwenk}},\ }\href
  {\doibase 10.1103/PhysRevLett.120.152503} {\bibfield  {journal} {\bibinfo
  {journal} {Phys. Rev. Lett.}\ }\textbf {\bibinfo {volume} {120}},\ \bibinfo
  {pages} {152503} (\bibinfo {year} {2018})}\BibitemShut {NoStop}%
\bibitem [{\citenamefont {Caurier}\ \emph {et~al.}(2005)\citenamefont
  {Caurier}, \citenamefont {Mart\'{\i}nez-Pinedo}, \citenamefont {Nowacki},
  \citenamefont {Poves},\ and\ \citenamefont {Zuker}}]{Caurier05a}%
  \BibitemOpen
  \bibfield  {author} {\bibinfo {author} {\bibfnamefont {E.}~\bibnamefont
  {Caurier}}, \bibinfo {author} {\bibfnamefont {G.}~\bibnamefont
  {Mart\'{\i}nez-Pinedo}}, \bibinfo {author} {\bibfnamefont {F.}~\bibnamefont
  {Nowacki}}, \bibinfo {author} {\bibfnamefont {A.}~\bibnamefont {Poves}}, \
  and\ \bibinfo {author} {\bibfnamefont {A.~P.}\ \bibnamefont {Zuker}},\ }\href
  {\doibase 10.1103/RevModPhys.77.427} {\bibfield  {journal} {\bibinfo
  {journal} {Rev. Mod. Phys.}\ }\textbf {\bibinfo {volume} {77}},\ \bibinfo
  {pages} {427} (\bibinfo {year} {2005})}\BibitemShut {NoStop}%
\bibitem [{\citenamefont {Otsuka}\ \emph {et~al.}(2001)\citenamefont {Otsuka},
  \citenamefont {Honma}, \citenamefont {Mizusaki}, \citenamefont {Shimizu},\
  and\ \citenamefont {Utsuno}}]{Otsuka01a}%
  \BibitemOpen
  \bibfield  {author} {\bibinfo {author} {\bibfnamefont {T.}~\bibnamefont
  {Otsuka}}, \bibinfo {author} {\bibfnamefont {M.}~\bibnamefont {Honma}},
  \bibinfo {author} {\bibfnamefont {T.}~\bibnamefont {Mizusaki}}, \bibinfo
  {author} {\bibfnamefont {N.}~\bibnamefont {Shimizu}}, \ and\ \bibinfo
  {author} {\bibfnamefont {Y.}~\bibnamefont {Utsuno}},\ }\href {\doibase
  https://doi.org/10.1016/S0146-6410(01)00157-0} {\bibfield  {journal}
  {\bibinfo  {journal} {Progress in Particle and Nuclear Physics}\ }\textbf
  {\bibinfo {volume} {47}},\ \bibinfo {pages} {319 } (\bibinfo {year}
  {2001})}\BibitemShut {NoStop}%
\bibitem [{\citenamefont {Bender}\ \emph {et~al.}(2003)\citenamefont {Bender},
  \citenamefont {Heenen},\ and\ \citenamefont {Reinhard}}]{Bender03a}%
  \BibitemOpen
  \bibfield  {author} {\bibinfo {author} {\bibfnamefont {M.}~\bibnamefont
  {Bender}}, \bibinfo {author} {\bibfnamefont {P.-H.}\ \bibnamefont {Heenen}},
  \ and\ \bibinfo {author} {\bibfnamefont {P.-G.}\ \bibnamefont {Reinhard}},\
  }\href {\doibase 10.1103/RevModPhys.75.121} {\bibfield  {journal} {\bibinfo
  {journal} {Rev. Mod. Phys.}\ }\textbf {\bibinfo {volume} {75}},\ \bibinfo
  {pages} {121} (\bibinfo {year} {2003})}\BibitemShut {NoStop}%
\bibitem [{\citenamefont {Nik\v{s}i\'c}\ \emph {et~al.}(2011)\citenamefont
  {Nik\v{s}i\'c}, \citenamefont {Vretenar},\ and\ \citenamefont
  {Ring}}]{Niksic11a}%
  \BibitemOpen
  \bibfield  {author} {\bibinfo {author} {\bibfnamefont {T.}~\bibnamefont
  {Nik\v{s}i\'c}}, \bibinfo {author} {\bibfnamefont {D.}~\bibnamefont
  {Vretenar}}, \ and\ \bibinfo {author} {\bibfnamefont {P.}~\bibnamefont
  {Ring}},\ }\href {\doibase https://doi.org/10.1016/j.ppnp.2011.01.055}
  {\bibfield  {journal} {\bibinfo  {journal} {Progress in Particle and Nuclear
  Physics}\ }\textbf {\bibinfo {volume} {66}},\ \bibinfo {pages} {519 }
  (\bibinfo {year} {2011})}\BibitemShut {NoStop}%
\bibitem [{\citenamefont {Egido}(2016)}]{Egido16a}%
  \BibitemOpen
  \bibfield  {author} {\bibinfo {author} {\bibfnamefont {J.~L.}\ \bibnamefont
  {Egido}},\ }\href {http://stacks.iop.org/1402-4896/91/i=7/a=073003}
  {\bibfield  {journal} {\bibinfo  {journal} {Physica Scripta}\ }\textbf
  {\bibinfo {volume} {91}},\ \bibinfo {pages} {073003} (\bibinfo {year}
  {2016})}\BibitemShut {NoStop}%
\bibitem [{\citenamefont {Robledo}\ \emph {et~al.}(2018)\citenamefont
  {Robledo}, \citenamefont {Rodr{\'{\i}}guez},\ and\ \citenamefont
  {Rodr{\'{\i}}guez-Guzm{\'{a}}n}}]{Robledo18a}%
  \BibitemOpen
  \bibfield  {author} {\bibinfo {author} {\bibfnamefont {L.~M.}\ \bibnamefont
  {Robledo}}, \bibinfo {author} {\bibfnamefont {T.~R.}\ \bibnamefont
  {Rodr{\'{\i}}guez}}, \ and\ \bibinfo {author} {\bibfnamefont {R.~R.}\
  \bibnamefont {Rodr{\'{\i}}guez-Guzm{\'{a}}n}},\ }\href {\doibase
  10.1088/1361-6471/aadebd} {\bibfield  {journal} {\bibinfo  {journal} {Journal
  of Physics G: Nuclear and Particle Physics}\ }\textbf {\bibinfo {volume}
  {46}},\ \bibinfo {pages} {013001} (\bibinfo {year} {2018})}\BibitemShut
  {NoStop}%
\bibitem [{\citenamefont {Duguet}(2014)}]{Duguet14b}%
  \BibitemOpen
  \bibfield  {author} {\bibinfo {author} {\bibfnamefont {T.}~\bibnamefont
  {Duguet}},\ }\enquote {\bibinfo {title} {The nuclear energy density
  functional formalism},}\ in\ \href {\doibase 10.1007/978-3-642-45141-6_7}
  {\emph {\bibinfo {booktitle} {The Euroschool on Exotic Beams, Vol. IV}}},\
  \bibinfo {editor} {edited by\ \bibinfo {editor} {\bibfnamefont
  {C.}~\bibnamefont {Scheidenberger}}\ and\ \bibinfo {editor} {\bibfnamefont
  {M.}~\bibnamefont {Pf{\"u}tzner}}}\ (\bibinfo  {publisher} {Springer Berlin
  Heidelberg},\ \bibinfo {address} {Berlin, Heidelberg},\ \bibinfo {year}
  {2014})\ pp.\ \bibinfo {pages} {293--350}\BibitemShut {NoStop}%
\bibitem [{\citenamefont {Jiao}\ \emph {et~al.}(2017)\citenamefont {Jiao},
  \citenamefont {Engel},\ and\ \citenamefont {Holt}}]{Jiao17a}%
  \BibitemOpen
  \bibfield  {author} {\bibinfo {author} {\bibfnamefont {C.~F.}\ \bibnamefont
  {Jiao}}, \bibinfo {author} {\bibfnamefont {J.}~\bibnamefont {Engel}}, \ and\
  \bibinfo {author} {\bibfnamefont {J.~D.}\ \bibnamefont {Holt}},\ }\href
  {\doibase 10.1103/PhysRevC.96.054310} {\bibfield  {journal} {\bibinfo
  {journal} {Phys. Rev. C}\ }\textbf {\bibinfo {volume} {96}},\ \bibinfo
  {pages} {054310} (\bibinfo {year} {2017})}\BibitemShut {NoStop}%
\bibitem [{\citenamefont {Yao}\ \emph {et~al.}(2018)\citenamefont {Yao},
  \citenamefont {Engel}, \citenamefont {Wang}, \citenamefont {Jiao},\ and\
  \citenamefont {Hergert}}]{Yao18a}%
  \BibitemOpen
  \bibfield  {author} {\bibinfo {author} {\bibfnamefont {J.~M.}\ \bibnamefont
  {Yao}}, \bibinfo {author} {\bibfnamefont {J.}~\bibnamefont {Engel}}, \bibinfo
  {author} {\bibfnamefont {L.~J.}\ \bibnamefont {Wang}}, \bibinfo {author}
  {\bibfnamefont {C.~F.}\ \bibnamefont {Jiao}}, \ and\ \bibinfo {author}
  {\bibfnamefont {H.}~\bibnamefont {Hergert}},\ }\href {\doibase
  10.1103/PhysRevC.98.054311} {\bibfield  {journal} {\bibinfo  {journal} {Phys.
  Rev. C}\ }\textbf {\bibinfo {volume} {98}},\ \bibinfo {pages} {054311}
  (\bibinfo {year} {2018})}\BibitemShut {NoStop}%
\bibitem [{\citenamefont {Jiao}\ \emph {et~al.}(2018)\citenamefont {Jiao},
  \citenamefont {Horoi},\ and\ \citenamefont {Neacsu}}]{Jiao18a}%
  \BibitemOpen
  \bibfield  {author} {\bibinfo {author} {\bibfnamefont {C.~F.}\ \bibnamefont
  {Jiao}}, \bibinfo {author} {\bibfnamefont {M.}~\bibnamefont {Horoi}}, \ and\
  \bibinfo {author} {\bibfnamefont {A.}~\bibnamefont {Neacsu}},\ }\href
  {\doibase 10.1103/PhysRevC.98.064324} {\bibfield  {journal} {\bibinfo
  {journal} {Phys. Rev. C}\ }\textbf {\bibinfo {volume} {98}},\ \bibinfo
  {pages} {064324} (\bibinfo {year} {2018})}\BibitemShut {NoStop}%
\bibitem [{\citenamefont {Maqbool}\ \emph {et~al.}(2011)\citenamefont
  {Maqbool}, \citenamefont {Sheikh}, \citenamefont {Ganai},\ and\ \citenamefont
  {Ring}}]{Maqbool11a}%
  \BibitemOpen
  \bibfield  {author} {\bibinfo {author} {\bibfnamefont {I.}~\bibnamefont
  {Maqbool}}, \bibinfo {author} {\bibfnamefont {J.~A.}\ \bibnamefont {Sheikh}},
  \bibinfo {author} {\bibfnamefont {P.~A.}\ \bibnamefont {Ganai}}, \ and\
  \bibinfo {author} {\bibfnamefont {P.}~\bibnamefont {Ring}},\ }\href {\doibase
  10.1088/0954-3899/38/4/045101} {\bibfield  {journal} {\bibinfo  {journal}
  {Journal of Physics G: Nuclear and Particle Physics}\ }\textbf {\bibinfo
  {volume} {38}},\ \bibinfo {pages} {045101} (\bibinfo {year}
  {2011})}\BibitemShut {NoStop}%
\bibitem [{\citenamefont {Hady\ifmmode \acute{n}\else
  \'{n}\fi{}ska-Kl\ifmmode~\mbox{\c{e}}\else \c{e}\fi{}k}\ \emph
  {et~al.}(2016)\citenamefont {Hady\ifmmode \acute{n}\else
  \'{n}\fi{}ska-Kl\ifmmode~\mbox{\c{e}}\else \c{e}\fi{}k}, \citenamefont
  {Napiorkowski}, \citenamefont {Zieli\ifmmode~\acute{n}\else \'{n}\fi{}ska},
  \citenamefont {Srebrny}, \citenamefont {Maj}, \citenamefont {Azaiez},
  \citenamefont {Valiente~Dob\'on}, \citenamefont {Kici\ifmmode
  \acute{n}\else~\'{n}\fi{}ska Habior}, \citenamefont {Nowacki}, \citenamefont
  {Na\"{\i}dja}, \citenamefont {Bounthong}, \citenamefont {Rodr\'{\i}guez},
  \citenamefont {de~Angelis}, \citenamefont {Abraham}, \citenamefont
  {Anil~Kumar}, \citenamefont {Bazzacco}, \citenamefont {Bellato},
  \citenamefont {Bortolato}, \citenamefont {Bednarczyk}, \citenamefont
  {Benzoni}, \citenamefont {Berti}, \citenamefont {Birkenbach}, \citenamefont
  {Bruyneel}, \citenamefont {Brambilla}, \citenamefont {Camera}, \citenamefont
  {Chavas}, \citenamefont {Cederwall}, \citenamefont {Charles}, \citenamefont
  {Ciema\l{}a}, \citenamefont {Cocconi}, \citenamefont {Coleman-Smith},
  \citenamefont {Colombo}, \citenamefont {Corsi}, \citenamefont {Crespi},
  \citenamefont {Cullen}, \citenamefont {Czermak}, \citenamefont
  {D\'esesquelles}, \citenamefont {Doherty}, \citenamefont {Dulny},
  \citenamefont {Eberth}, \citenamefont {Farnea}, \citenamefont {Fornal},
  \citenamefont {Franchoo}, \citenamefont {Gadea}, \citenamefont {Giaz},
  \citenamefont {Gottardo}, \citenamefont {Grave}, \citenamefont
  {Gr\ifmmode~\mbox{\c{e}}\else \c{e}\fi{}bosz}, \citenamefont {G\"orgen},
  \citenamefont {Gulmini}, \citenamefont {Habermann}, \citenamefont {Hess},
  \citenamefont {Isocrate}, \citenamefont {Iwanicki}, \citenamefont {Jaworski},
  \citenamefont {Judson}, \citenamefont {Jungclaus}, \citenamefont {Karkour},
  \citenamefont {Kmiecik}, \citenamefont {Karpi\ifmmode~\acute{n}\else
  \'{n}\fi{}ski}, \citenamefont {Kisieli\ifmmode~\acute{n}\else \'{n}\fi{}ski},
  \citenamefont {Kondratyev}, \citenamefont {Korichi}, \citenamefont
  {Komorowska}, \citenamefont {Kowalczyk}, \citenamefont {Korten},
  \citenamefont {Krzysiek}, \citenamefont {Lehaut}, \citenamefont {Leoni},
  \citenamefont {Ljungvall}, \citenamefont {Lopez-Martens}, \citenamefont
  {Lunardi}, \citenamefont {Maron}, \citenamefont {Mazurek}, \citenamefont
  {Menegazzo}, \citenamefont {Mengoni}, \citenamefont {Merch\'an},
  \citenamefont {M\ifmmode \mbox{\c{e}}\else
  \c{e}\fi{}czy\ifmmode~\acute{n}\else \'{n}\fi{}ski}, \citenamefont
  {Michelagnoli}, \citenamefont {Mierzejewski}, \citenamefont {Million},
  \citenamefont {Myalski}, \citenamefont {Napoli}, \citenamefont {Nicolini},
  \citenamefont {Niikura}, \citenamefont {Obertelli}, \citenamefont {\"Ozmen},
  \citenamefont {Palacz}, \citenamefont {Pr\'ochniak}, \citenamefont {Pullia},
  \citenamefont {Quintana}, \citenamefont {Rampazzo}, \citenamefont {Recchia},
  \citenamefont {Redon}, \citenamefont {Reiter}, \citenamefont {Rosso},
  \citenamefont {Rusek}, \citenamefont {Sahin}, \citenamefont {Salsac},
  \citenamefont {S\"oderstr\"om}, \citenamefont {Stefan}, \citenamefont
  {St\'ezowski}, \citenamefont {Stycze\ifmmode~\acute{n}\else \'{n}\fi{}},
  \citenamefont {Theisen}, \citenamefont {Toniolo}, \citenamefont {Ur},
  \citenamefont {Vandone}, \citenamefont {Wadsworth}, \citenamefont
  {Wasilewska}, \citenamefont {Wiens}, \citenamefont {Wood}, \citenamefont
  {Wrzosek-Lipska},\ and\ \citenamefont {Zi\ifmmode \mbox{\c{e}}\else
  \c{e}\fi{}bli\ifmmode~\acute{n}\else \'{n}\fi{}ski}}]{Hadynska16a}%
  \BibitemOpen
  \bibfield  {author} {\bibinfo {author} {\bibfnamefont {K.}~\bibnamefont
  {Hady\ifmmode \acute{n}\else \'{n}\fi{}ska-Kl\ifmmode~\mbox{\c{e}}\else
  \c{e}\fi{}k}}, \bibinfo {author} {\bibfnamefont {P.~J.}\ \bibnamefont
  {Napiorkowski}}, \bibinfo {author} {\bibfnamefont {M.}~\bibnamefont
  {Zieli\ifmmode~\acute{n}\else \'{n}\fi{}ska}}, \bibinfo {author}
  {\bibfnamefont {J.}~\bibnamefont {Srebrny}}, \bibinfo {author} {\bibfnamefont
  {A.}~\bibnamefont {Maj}}, \bibinfo {author} {\bibfnamefont {F.}~\bibnamefont
  {Azaiez}}, \bibinfo {author} {\bibfnamefont {J.~J.}\ \bibnamefont
  {Valiente~Dob\'on}}, \bibinfo {author} {\bibfnamefont {M.}~\bibnamefont
  {Kici\ifmmode \acute{n}\else~\'{n}\fi{}ska Habior}}, \bibinfo {author}
  {\bibfnamefont {F.}~\bibnamefont {Nowacki}}, \bibinfo {author} {\bibfnamefont
  {H.}~\bibnamefont {Na\"{\i}dja}}, \bibinfo {author} {\bibfnamefont
  {B.}~\bibnamefont {Bounthong}}, \bibinfo {author} {\bibfnamefont {T.~R.}\
  \bibnamefont {Rodr\'{\i}guez}}, \bibinfo {author} {\bibfnamefont
  {G.}~\bibnamefont {de~Angelis}}, \bibinfo {author} {\bibfnamefont
  {T.}~\bibnamefont {Abraham}}, \bibinfo {author} {\bibfnamefont
  {G.}~\bibnamefont {Anil~Kumar}}, \bibinfo {author} {\bibfnamefont
  {D.}~\bibnamefont {Bazzacco}}, \bibinfo {author} {\bibfnamefont
  {M.}~\bibnamefont {Bellato}}, \bibinfo {author} {\bibfnamefont
  {D.}~\bibnamefont {Bortolato}}, \bibinfo {author} {\bibfnamefont
  {P.}~\bibnamefont {Bednarczyk}}, \bibinfo {author} {\bibfnamefont
  {G.}~\bibnamefont {Benzoni}}, \bibinfo {author} {\bibfnamefont
  {L.}~\bibnamefont {Berti}}, \bibinfo {author} {\bibfnamefont
  {B.}~\bibnamefont {Birkenbach}}, \bibinfo {author} {\bibfnamefont
  {B.}~\bibnamefont {Bruyneel}}, \bibinfo {author} {\bibfnamefont
  {S.}~\bibnamefont {Brambilla}}, \bibinfo {author} {\bibfnamefont
  {F.}~\bibnamefont {Camera}}, \bibinfo {author} {\bibfnamefont
  {J.}~\bibnamefont {Chavas}}, \bibinfo {author} {\bibfnamefont
  {B.}~\bibnamefont {Cederwall}}, \bibinfo {author} {\bibfnamefont
  {L.}~\bibnamefont {Charles}}, \bibinfo {author} {\bibfnamefont
  {M.}~\bibnamefont {Ciema\l{}a}}, \bibinfo {author} {\bibfnamefont
  {P.}~\bibnamefont {Cocconi}}, \bibinfo {author} {\bibfnamefont
  {P.}~\bibnamefont {Coleman-Smith}}, \bibinfo {author} {\bibfnamefont
  {A.}~\bibnamefont {Colombo}}, \bibinfo {author} {\bibfnamefont
  {A.}~\bibnamefont {Corsi}}, \bibinfo {author} {\bibfnamefont {F.~C.~L.}\
  \bibnamefont {Crespi}}, \bibinfo {author} {\bibfnamefont {D.~M.}\
  \bibnamefont {Cullen}}, \bibinfo {author} {\bibfnamefont {A.}~\bibnamefont
  {Czermak}}, \bibinfo {author} {\bibfnamefont {P.}~\bibnamefont
  {D\'esesquelles}}, \bibinfo {author} {\bibfnamefont {D.~T.}\ \bibnamefont
  {Doherty}}, \bibinfo {author} {\bibfnamefont {B.}~\bibnamefont {Dulny}},
  \bibinfo {author} {\bibfnamefont {J.}~\bibnamefont {Eberth}}, \bibinfo
  {author} {\bibfnamefont {E.}~\bibnamefont {Farnea}}, \bibinfo {author}
  {\bibfnamefont {B.}~\bibnamefont {Fornal}}, \bibinfo {author} {\bibfnamefont
  {S.}~\bibnamefont {Franchoo}}, \bibinfo {author} {\bibfnamefont
  {A.}~\bibnamefont {Gadea}}, \bibinfo {author} {\bibfnamefont
  {A.}~\bibnamefont {Giaz}}, \bibinfo {author} {\bibfnamefont {A.}~\bibnamefont
  {Gottardo}}, \bibinfo {author} {\bibfnamefont {X.}~\bibnamefont {Grave}},
  \bibinfo {author} {\bibfnamefont {J.}~\bibnamefont
  {Gr\ifmmode~\mbox{\c{e}}\else \c{e}\fi{}bosz}}, \bibinfo {author}
  {\bibfnamefont {A.}~\bibnamefont {G\"orgen}}, \bibinfo {author}
  {\bibfnamefont {M.}~\bibnamefont {Gulmini}}, \bibinfo {author} {\bibfnamefont
  {T.}~\bibnamefont {Habermann}}, \bibinfo {author} {\bibfnamefont
  {H.}~\bibnamefont {Hess}}, \bibinfo {author} {\bibfnamefont {R.}~\bibnamefont
  {Isocrate}}, \bibinfo {author} {\bibfnamefont {J.}~\bibnamefont {Iwanicki}},
  \bibinfo {author} {\bibfnamefont {G.}~\bibnamefont {Jaworski}}, \bibinfo
  {author} {\bibfnamefont {D.~S.}\ \bibnamefont {Judson}}, \bibinfo {author}
  {\bibfnamefont {A.}~\bibnamefont {Jungclaus}}, \bibinfo {author}
  {\bibfnamefont {N.}~\bibnamefont {Karkour}}, \bibinfo {author} {\bibfnamefont
  {M.}~\bibnamefont {Kmiecik}}, \bibinfo {author} {\bibfnamefont
  {D.}~\bibnamefont {Karpi\ifmmode~\acute{n}\else \'{n}\fi{}ski}}, \bibinfo
  {author} {\bibfnamefont {M.}~\bibnamefont {Kisieli\ifmmode~\acute{n}\else
  \'{n}\fi{}ski}}, \bibinfo {author} {\bibfnamefont {N.}~\bibnamefont
  {Kondratyev}}, \bibinfo {author} {\bibfnamefont {A.}~\bibnamefont {Korichi}},
  \bibinfo {author} {\bibfnamefont {M.}~\bibnamefont {Komorowska}}, \bibinfo
  {author} {\bibfnamefont {M.}~\bibnamefont {Kowalczyk}}, \bibinfo {author}
  {\bibfnamefont {W.}~\bibnamefont {Korten}}, \bibinfo {author} {\bibfnamefont
  {M.}~\bibnamefont {Krzysiek}}, \bibinfo {author} {\bibfnamefont
  {G.}~\bibnamefont {Lehaut}}, \bibinfo {author} {\bibfnamefont
  {S.}~\bibnamefont {Leoni}}, \bibinfo {author} {\bibfnamefont
  {J.}~\bibnamefont {Ljungvall}}, \bibinfo {author} {\bibfnamefont
  {A.}~\bibnamefont {Lopez-Martens}}, \bibinfo {author} {\bibfnamefont
  {S.}~\bibnamefont {Lunardi}}, \bibinfo {author} {\bibfnamefont
  {G.}~\bibnamefont {Maron}}, \bibinfo {author} {\bibfnamefont
  {K.}~\bibnamefont {Mazurek}}, \bibinfo {author} {\bibfnamefont
  {R.}~\bibnamefont {Menegazzo}}, \bibinfo {author} {\bibfnamefont
  {D.}~\bibnamefont {Mengoni}}, \bibinfo {author} {\bibfnamefont
  {E.}~\bibnamefont {Merch\'an}}, \bibinfo {author} {\bibfnamefont
  {W.}~\bibnamefont {M\ifmmode \mbox{\c{e}}\else
  \c{e}\fi{}czy\ifmmode~\acute{n}\else \'{n}\fi{}ski}}, \bibinfo {author}
  {\bibfnamefont {C.}~\bibnamefont {Michelagnoli}}, \bibinfo {author}
  {\bibfnamefont {J.}~\bibnamefont {Mierzejewski}}, \bibinfo {author}
  {\bibfnamefont {B.}~\bibnamefont {Million}}, \bibinfo {author} {\bibfnamefont
  {S.}~\bibnamefont {Myalski}}, \bibinfo {author} {\bibfnamefont {D.~R.}\
  \bibnamefont {Napoli}}, \bibinfo {author} {\bibfnamefont {R.}~\bibnamefont
  {Nicolini}}, \bibinfo {author} {\bibfnamefont {M.}~\bibnamefont {Niikura}},
  \bibinfo {author} {\bibfnamefont {A.}~\bibnamefont {Obertelli}}, \bibinfo
  {author} {\bibfnamefont {S.~F.}\ \bibnamefont {\"Ozmen}}, \bibinfo {author}
  {\bibfnamefont {M.}~\bibnamefont {Palacz}}, \bibinfo {author} {\bibfnamefont
  {L.}~\bibnamefont {Pr\'ochniak}}, \bibinfo {author} {\bibfnamefont
  {A.}~\bibnamefont {Pullia}}, \bibinfo {author} {\bibfnamefont
  {B.}~\bibnamefont {Quintana}}, \bibinfo {author} {\bibfnamefont
  {G.}~\bibnamefont {Rampazzo}}, \bibinfo {author} {\bibfnamefont
  {F.}~\bibnamefont {Recchia}}, \bibinfo {author} {\bibfnamefont
  {N.}~\bibnamefont {Redon}}, \bibinfo {author} {\bibfnamefont
  {P.}~\bibnamefont {Reiter}}, \bibinfo {author} {\bibfnamefont
  {D.}~\bibnamefont {Rosso}}, \bibinfo {author} {\bibfnamefont
  {K.}~\bibnamefont {Rusek}}, \bibinfo {author} {\bibfnamefont
  {E.}~\bibnamefont {Sahin}}, \bibinfo {author} {\bibfnamefont {M.-D.}\
  \bibnamefont {Salsac}}, \bibinfo {author} {\bibfnamefont {P.-A.}\
  \bibnamefont {S\"oderstr\"om}}, \bibinfo {author} {\bibfnamefont
  {I.}~\bibnamefont {Stefan}}, \bibinfo {author} {\bibfnamefont
  {O.}~\bibnamefont {St\'ezowski}}, \bibinfo {author} {\bibfnamefont
  {J.}~\bibnamefont {Stycze\ifmmode~\acute{n}\else \'{n}\fi{}}}, \bibinfo
  {author} {\bibfnamefont {C.}~\bibnamefont {Theisen}}, \bibinfo {author}
  {\bibfnamefont {N.}~\bibnamefont {Toniolo}}, \bibinfo {author} {\bibfnamefont
  {C.~A.}\ \bibnamefont {Ur}}, \bibinfo {author} {\bibfnamefont
  {V.}~\bibnamefont {Vandone}}, \bibinfo {author} {\bibfnamefont
  {R.}~\bibnamefont {Wadsworth}}, \bibinfo {author} {\bibfnamefont
  {B.}~\bibnamefont {Wasilewska}}, \bibinfo {author} {\bibfnamefont
  {A.}~\bibnamefont {Wiens}}, \bibinfo {author} {\bibfnamefont {J.~L.}\
  \bibnamefont {Wood}}, \bibinfo {author} {\bibfnamefont {K.}~\bibnamefont
  {Wrzosek-Lipska}}, \ and\ \bibinfo {author} {\bibfnamefont {M.}~\bibnamefont
  {Zi\ifmmode \mbox{\c{e}}\else \c{e}\fi{}bli\ifmmode~\acute{n}\else
  \'{n}\fi{}ski}},\ }\href {\doibase 10.1103/PhysRevLett.117.062501} {\bibfield
   {journal} {\bibinfo  {journal} {Phys. Rev. Lett.}\ }\textbf {\bibinfo
  {volume} {117}},\ \bibinfo {pages} {062501} (\bibinfo {year}
  {2016})}\BibitemShut {NoStop}%
\bibitem [{\citenamefont {Bounthong}(2016)}]{BounthongPHD}%
  \BibitemOpen
  \bibfield  {author} {\bibinfo {author} {\bibfnamefont {B.}~\bibnamefont
  {Bounthong}},\ }\href@noop {} {Ph.D. thesis},\ \bibinfo  {school}
  {Universit{\'e} de Strasbourg} (\bibinfo {year} {2016})\BibitemShut {NoStop}%
\bibitem [{\citenamefont {Tsunoda}\ \emph {et~al.}(2014)\citenamefont
  {Tsunoda}, \citenamefont {Otsuka}, \citenamefont {Shimizu}, \citenamefont
  {Honma},\ and\ \citenamefont {Utsuno}}]{Tsunoda14a}%
  \BibitemOpen
  \bibfield  {author} {\bibinfo {author} {\bibfnamefont {Y.}~\bibnamefont
  {Tsunoda}}, \bibinfo {author} {\bibfnamefont {T.}~\bibnamefont {Otsuka}},
  \bibinfo {author} {\bibfnamefont {N.}~\bibnamefont {Shimizu}}, \bibinfo
  {author} {\bibfnamefont {M.}~\bibnamefont {Honma}}, \ and\ \bibinfo {author}
  {\bibfnamefont {Y.}~\bibnamefont {Utsuno}},\ }\href {\doibase
  10.1103/PhysRevC.89.031301} {\bibfield  {journal} {\bibinfo  {journal} {Phys.
  Rev. C}\ }\textbf {\bibinfo {volume} {89}},\ \bibinfo {pages} {031301}
  (\bibinfo {year} {2014})}\BibitemShut {NoStop}%
\bibitem [{\citenamefont {Togashi}\ \emph {et~al.}(2018)\citenamefont
  {Togashi}, \citenamefont {Tsunoda}, \citenamefont {Otsuka}, \citenamefont
  {Shimizu},\ and\ \citenamefont {Honma}}]{Togashi18a}%
  \BibitemOpen
  \bibfield  {author} {\bibinfo {author} {\bibfnamefont {T.}~\bibnamefont
  {Togashi}}, \bibinfo {author} {\bibfnamefont {Y.}~\bibnamefont {Tsunoda}},
  \bibinfo {author} {\bibfnamefont {T.}~\bibnamefont {Otsuka}}, \bibinfo
  {author} {\bibfnamefont {N.}~\bibnamefont {Shimizu}}, \ and\ \bibinfo
  {author} {\bibfnamefont {M.}~\bibnamefont {Honma}},\ }\href {\doibase
  10.1103/PhysRevLett.121.062501} {\bibfield  {journal} {\bibinfo  {journal}
  {Phys. Rev. Lett.}\ }\textbf {\bibinfo {volume} {121}},\ \bibinfo {pages}
  {062501} (\bibinfo {year} {2018})}\BibitemShut {NoStop}%
\bibitem [{\citenamefont {Hara}\ and\ \citenamefont {Sun}(1995)}]{Hara95a}%
  \BibitemOpen
  \bibfield  {author} {\bibinfo {author} {\bibfnamefont {K.}~\bibnamefont
  {Hara}}\ and\ \bibinfo {author} {\bibfnamefont {Y.}~\bibnamefont {Sun}},\
  }\href {\doibase 10.1142/S0218301395000250} {\bibfield  {journal} {\bibinfo
  {journal} {International Journal of Modern Physics E}\ }\textbf {\bibinfo
  {volume} {04}},\ \bibinfo {pages} {637} (\bibinfo {year} {1995})},\ \Eprint
  {http://arxiv.org/abs/https://doi.org/10.1142/S0218301395000250}
  {https://doi.org/10.1142/S0218301395000250} \BibitemShut {NoStop}%
\bibitem [{\citenamefont {Sun}(2016)}]{Sun16a}%
  \BibitemOpen
  \bibfield  {author} {\bibinfo {author} {\bibfnamefont {Y.}~\bibnamefont
  {Sun}},\ }\href {\doibase 10.1088/0031-8949/91/4/043005} {\bibfield
  {journal} {\bibinfo  {journal} {Physica Scripta}\ }\textbf {\bibinfo {volume}
  {91}},\ \bibinfo {pages} {043005} (\bibinfo {year} {2016})}\BibitemShut
  {NoStop}%
\bibitem [{\citenamefont {Chen}\ and\ \citenamefont {Egido}(2016)}]{Chen16a}%
  \BibitemOpen
  \bibfield  {author} {\bibinfo {author} {\bibfnamefont {F.-Q.}\ \bibnamefont
  {Chen}}\ and\ \bibinfo {author} {\bibfnamefont {J.~L.}\ \bibnamefont
  {Egido}},\ }\href {\doibase 10.1103/PhysRevC.93.064313} {\bibfield  {journal}
  {\bibinfo  {journal} {Phys. Rev. C}\ }\textbf {\bibinfo {volume} {93}},\
  \bibinfo {pages} {064313} (\bibinfo {year} {2016})}\BibitemShut {NoStop}%
\bibitem [{\citenamefont {Chen}\ and\ \citenamefont {Egido}(2017)}]{Chen17a}%
  \BibitemOpen
  \bibfield  {author} {\bibinfo {author} {\bibfnamefont {F.-Q.}\ \bibnamefont
  {Chen}}\ and\ \bibinfo {author} {\bibfnamefont {J.~L.}\ \bibnamefont
  {Egido}},\ }\href {\doibase 10.1103/PhysRevC.95.024307} {\bibfield  {journal}
  {\bibinfo  {journal} {Phys. Rev. C}\ }\textbf {\bibinfo {volume} {95}},\
  \bibinfo {pages} {024307} (\bibinfo {year} {2017})}\BibitemShut {NoStop}%
\bibitem [{\citenamefont {Schmid}\ and\ \citenamefont
  {Grummer}(1987)}]{Schmid87a}%
  \BibitemOpen
  \bibfield  {author} {\bibinfo {author} {\bibfnamefont {K.~W.}\ \bibnamefont
  {Schmid}}\ and\ \bibinfo {author} {\bibfnamefont {F.}~\bibnamefont
  {Grummer}},\ }\href {\doibase 10.1088/0034-4885/50/6/003} {\bibfield
  {journal} {\bibinfo  {journal} {Reports on Progress in Physics}\ }\textbf
  {\bibinfo {volume} {50}},\ \bibinfo {pages} {731} (\bibinfo {year}
  {1987})}\BibitemShut {NoStop}%
\bibitem [{\citenamefont {Schmid}(2004)}]{Schmid04a}%
  \BibitemOpen
  \bibfield  {author} {\bibinfo {author} {\bibfnamefont {K.~W.}\ \bibnamefont
  {Schmid}},\ }\href@noop {} {\bibfield  {journal} {\bibinfo  {journal} {Prog.
  Part. Nucl. Phys.}\ }\textbf {\bibinfo {volume} {52}},\ \bibinfo {pages}
  {565} (\bibinfo {year} {2004})}\BibitemShut {NoStop}%
\bibitem [{\citenamefont {Poves}\ \emph {et~al.}(2001)\citenamefont {Poves},
  \citenamefont {Sánchez-Solano}, \citenamefont {Caurier},\ and\ \citenamefont
  {Nowacki}}]{Poves01a}%
  \BibitemOpen
  \bibfield  {author} {\bibinfo {author} {\bibfnamefont {A.}~\bibnamefont
  {Poves}}, \bibinfo {author} {\bibfnamefont {J.}~\bibnamefont
  {Sánchez-Solano}}, \bibinfo {author} {\bibfnamefont {E.}~\bibnamefont
  {Caurier}}, \ and\ \bibinfo {author} {\bibfnamefont {F.}~\bibnamefont
  {Nowacki}},\ }\href {\doibase https://doi.org/10.1016/S0375-9474(01)00967-8}
  {\bibfield  {journal} {\bibinfo  {journal} {Nuclear Physics A}\ }\textbf
  {\bibinfo {volume} {694}},\ \bibinfo {pages} {157 } (\bibinfo {year}
  {2001})}\BibitemShut {NoStop}%
\bibitem [{\citenamefont {Ring}\ and\ \citenamefont {Schuck}(1980)}]{RS80a}%
  \BibitemOpen
  \bibfield  {author} {\bibinfo {author} {\bibfnamefont {P.}~\bibnamefont
  {Ring}}\ and\ \bibinfo {author} {\bibfnamefont {P.}~\bibnamefont {Schuck}},\
  }\href@noop {} {\emph {\bibinfo {title} {The Nuclear Many-Body Problem}}}\
  (\bibinfo  {publisher} {Springer-Verlag},\ \bibinfo {address} {New York},\
  \bibinfo {year} {1980})\BibitemShut {NoStop}%
\bibitem [{\citenamefont {Bally}(2014)}]{BallyPHD}%
  \BibitemOpen
  \bibfield  {author} {\bibinfo {author} {\bibfnamefont {B.}~\bibnamefont
  {Bally}},\ }\emph {\bibinfo {title} {Description of odd-mass nuclei by
  multi-reference energy density functional methods}},\ \href@noop {} {Ph.D.
  thesis},\ \bibinfo  {school} {Universit{\'e} de Bordeaux} (\bibinfo {year}
  {2014})\BibitemShut {NoStop}%
\bibitem [{\citenamefont {Anguiano}\ \emph {et~al.}(2001)\citenamefont
  {Anguiano}, \citenamefont {Egido},\ and\ \citenamefont
  {Robledo}}]{Anguiano01a}%
  \BibitemOpen
  \bibfield  {author} {\bibinfo {author} {\bibfnamefont {M.}~\bibnamefont
  {Anguiano}}, \bibinfo {author} {\bibfnamefont {J.}~\bibnamefont {Egido}}, \
  and\ \bibinfo {author} {\bibfnamefont {L.}~\bibnamefont {Robledo}},\ }\href
  {\doibase https://doi.org/10.1016/S0375-9474(01)01219-2} {\bibfield
  {journal} {\bibinfo  {journal} {Nuclear Physics A}\ }\textbf {\bibinfo
  {volume} {696}},\ \bibinfo {pages} {467 } (\bibinfo {year}
  {2001})}\BibitemShut {NoStop}%
\bibitem [{\citenamefont {Lacroix}\ \emph {et~al.}(2009)\citenamefont
  {Lacroix}, \citenamefont {Duguet},\ and\ \citenamefont
  {Bender}}]{Lacroix09a}%
  \BibitemOpen
  \bibfield  {author} {\bibinfo {author} {\bibfnamefont {D.}~\bibnamefont
  {Lacroix}}, \bibinfo {author} {\bibfnamefont {T.}~\bibnamefont {Duguet}}, \
  and\ \bibinfo {author} {\bibfnamefont {M.}~\bibnamefont {Bender}},\ }\href
  {\doibase 10.1103/PhysRevC.79.044318} {\bibfield  {journal} {\bibinfo
  {journal} {Phys. Rev. C}\ }\textbf {\bibinfo {volume} {79}},\ \bibinfo
  {pages} {044318} (\bibinfo {year} {2009})}\BibitemShut {NoStop}%
\bibitem [{\citenamefont {Bender}\ \emph {et~al.}(2009)\citenamefont {Bender},
  \citenamefont {Duguet},\ and\ \citenamefont {Lacroix}}]{Bender09a}%
  \BibitemOpen
  \bibfield  {author} {\bibinfo {author} {\bibfnamefont {M.}~\bibnamefont
  {Bender}}, \bibinfo {author} {\bibfnamefont {T.}~\bibnamefont {Duguet}}, \
  and\ \bibinfo {author} {\bibfnamefont {D.}~\bibnamefont {Lacroix}},\ }\href
  {\doibase 10.1103/PhysRevC.79.044319} {\bibfield  {journal} {\bibinfo
  {journal} {Phys. Rev. C}\ }\textbf {\bibinfo {volume} {79}},\ \bibinfo
  {pages} {044319} (\bibinfo {year} {2009})}\BibitemShut {NoStop}%
\bibitem [{\citenamefont {Duguet}\ \emph {et~al.}(2009)\citenamefont {Duguet},
  \citenamefont {Bender}, \citenamefont {Bennaceur}, \citenamefont {Lacroix},\
  and\ \citenamefont {Lesinski}}]{Duguet09a}%
  \BibitemOpen
  \bibfield  {author} {\bibinfo {author} {\bibfnamefont {T.}~\bibnamefont
  {Duguet}}, \bibinfo {author} {\bibfnamefont {M.}~\bibnamefont {Bender}},
  \bibinfo {author} {\bibfnamefont {K.}~\bibnamefont {Bennaceur}}, \bibinfo
  {author} {\bibfnamefont {D.}~\bibnamefont {Lacroix}}, \ and\ \bibinfo
  {author} {\bibfnamefont {T.}~\bibnamefont {Lesinski}},\ }\href {\doibase
  10.1103/PhysRevC.79.044320} {\bibfield  {journal} {\bibinfo  {journal} {Phys.
  Rev. C}\ }\textbf {\bibinfo {volume} {79}},\ \bibinfo {pages} {044320}
  (\bibinfo {year} {2009})}\BibitemShut {NoStop}%
\bibitem [{\citenamefont {Robledo}(2009)}]{Robledo09a}%
  \BibitemOpen
  \bibfield  {author} {\bibinfo {author} {\bibfnamefont {L.~M.}\ \bibnamefont
  {Robledo}},\ }\href {\doibase 10.1103/PhysRevC.79.021302} {\bibfield
  {journal} {\bibinfo  {journal} {Phys. Rev. C}\ }\textbf {\bibinfo {volume}
  {79}},\ \bibinfo {pages} {021302} (\bibinfo {year} {2009})}\BibitemShut
  {NoStop}%
\bibitem [{\citenamefont {Avez}\ and\ \citenamefont {Bender}(2012)}]{Avez12a}%
  \BibitemOpen
  \bibfield  {author} {\bibinfo {author} {\bibfnamefont {B.}~\bibnamefont
  {Avez}}\ and\ \bibinfo {author} {\bibfnamefont {M.}~\bibnamefont {Bender}},\
  }\href {\doibase 10.1103/PhysRevC.85.034325} {\bibfield  {journal} {\bibinfo
  {journal} {Phys. Rev. C}\ }\textbf {\bibinfo {volume} {85}},\ \bibinfo
  {pages} {034325} (\bibinfo {year} {2012})}\BibitemShut {NoStop}%
\bibitem [{\citenamefont {Wimmer}(2012)}]{Wimmer12a}%
  \BibitemOpen
  \bibfield  {author} {\bibinfo {author} {\bibfnamefont {M.}~\bibnamefont
  {Wimmer}},\ }\href {\doibase 10.1145/2331130.2331138} {\bibfield  {journal}
  {\bibinfo  {journal} {ACM Trans. Math. Softw.}\ }\textbf {\bibinfo {volume}
  {38}},\ \bibinfo {pages} {30:1} (\bibinfo {year} {2012})}\BibitemShut
  {NoStop}%
\bibitem [{\citenamefont {Bally}\ and\ \citenamefont
  {Duguet}(2018)}]{Bally18a}%
  \BibitemOpen
  \bibfield  {author} {\bibinfo {author} {\bibfnamefont {B.}~\bibnamefont
  {Bally}}\ and\ \bibinfo {author} {\bibfnamefont {T.}~\bibnamefont {Duguet}},\
  }\href {\doibase 10.1103/PhysRevC.97.024304} {\bibfield  {journal} {\bibinfo
  {journal} {Phys. Rev. C}\ }\textbf {\bibinfo {volume} {97}},\ \bibinfo
  {pages} {024304} (\bibinfo {year} {2018})}\BibitemShut {NoStop}%
\bibitem [{\citenamefont {Mizusaki}\ \emph {et~al.}(2018)\citenamefont
  {Mizusaki}, \citenamefont {Oi},\ and\ \citenamefont {Shimizu}}]{Mizusaki18a}%
  \BibitemOpen
  \bibfield  {author} {\bibinfo {author} {\bibfnamefont {T.}~\bibnamefont
  {Mizusaki}}, \bibinfo {author} {\bibfnamefont {M.}~\bibnamefont {Oi}}, \ and\
  \bibinfo {author} {\bibfnamefont {N.}~\bibnamefont {Shimizu}},\ }\href
  {\doibase https://doi.org/10.1016/j.physletb.2018.02.012} {\bibfield
  {journal} {\bibinfo  {journal} {Physics Letters B}\ }\textbf {\bibinfo
  {volume} {779}},\ \bibinfo {pages} {237 } (\bibinfo {year}
  {2018})}\BibitemShut {NoStop}%
\bibitem [{\citenamefont {Rodr\'{\i}guez}\ \emph {et~al.}(2005)\citenamefont
  {Rodr\'{\i}guez}, \citenamefont {Egido},\ and\ \citenamefont
  {Robledo}}]{Rodriguez05a}%
  \BibitemOpen
  \bibfield  {author} {\bibinfo {author} {\bibfnamefont {T.~R.}\ \bibnamefont
  {Rodr\'{\i}guez}}, \bibinfo {author} {\bibfnamefont {J.~L.}\ \bibnamefont
  {Egido}}, \ and\ \bibinfo {author} {\bibfnamefont {L.~M.}\ \bibnamefont
  {Robledo}},\ }\href {\doibase 10.1103/PhysRevC.72.064303} {\bibfield
  {journal} {\bibinfo  {journal} {Phys. Rev. C}\ }\textbf {\bibinfo {volume}
  {72}},\ \bibinfo {pages} {064303} (\bibinfo {year} {2005})}\BibitemShut
  {NoStop}%
\bibitem [{\citenamefont {Rodr\'{\i}guez}\ and\ \citenamefont
  {Egido}(2007)}]{Rodriguez07a}%
  \BibitemOpen
  \bibfield  {author} {\bibinfo {author} {\bibfnamefont {T.~R.}\ \bibnamefont
  {Rodr\'{\i}guez}}\ and\ \bibinfo {author} {\bibfnamefont {J.~L.}\
  \bibnamefont {Egido}},\ }\href {\doibase 10.1103/PhysRevLett.99.062501}
  {\bibfield  {journal} {\bibinfo  {journal} {Phys. Rev. Lett.}\ }\textbf
  {\bibinfo {volume} {99}},\ \bibinfo {pages} {062501} (\bibinfo {year}
  {2007})}\BibitemShut {NoStop}%
\bibitem [{\citenamefont {Robledo}\ and\ \citenamefont
  {Bertsch}(2011)}]{Robledo11a}%
  \BibitemOpen
  \bibfield  {author} {\bibinfo {author} {\bibfnamefont {L.~M.}\ \bibnamefont
  {Robledo}}\ and\ \bibinfo {author} {\bibfnamefont {G.~F.}\ \bibnamefont
  {Bertsch}},\ }\href {\doibase 10.1103/PhysRevC.84.014312} {\bibfield
  {journal} {\bibinfo  {journal} {Phys. Rev. C}\ }\textbf {\bibinfo {volume}
  {84}},\ \bibinfo {pages} {014312} (\bibinfo {year} {2011})}\BibitemShut
  {NoStop}%
\bibitem [{\citenamefont {Egido}\ \emph {et~al.}(1995)\citenamefont {Egido},
  \citenamefont {Lessing}, \citenamefont {Martin},\ and\ \citenamefont
  {Robledo}}]{Egido95a}%
  \BibitemOpen
  \bibfield  {author} {\bibinfo {author} {\bibfnamefont {J.}~\bibnamefont
  {Egido}}, \bibinfo {author} {\bibfnamefont {J.}~\bibnamefont {Lessing}},
  \bibinfo {author} {\bibfnamefont {V.}~\bibnamefont {Martin}}, \ and\ \bibinfo
  {author} {\bibfnamefont {L.}~\bibnamefont {Robledo}},\ }\href {\doibase
  https://doi.org/10.1016/0375-9474(95)00370-G} {\bibfield  {journal} {\bibinfo
   {journal} {Nuclear Physics A}\ }\textbf {\bibinfo {volume} {594}},\ \bibinfo
  {pages} {70 } (\bibinfo {year} {1995})}\BibitemShut {NoStop}%
\bibitem [{\citenamefont {Ryssens}\ \emph {et~al.}(2019)\citenamefont
  {Ryssens}, \citenamefont {Bender},\ and\ \citenamefont
  {Heenen}}]{Ryssens19a}%
  \BibitemOpen
  \bibfield  {author} {\bibinfo {author} {\bibfnamefont {W.}~\bibnamefont
  {Ryssens}}, \bibinfo {author} {\bibfnamefont {M.}~\bibnamefont {Bender}}, \
  and\ \bibinfo {author} {\bibfnamefont {P.~H.}\ \bibnamefont {Heenen}},\
  }\href {\doibase 10.1140/epja/i2019-12766-6} {\bibfield  {journal} {\bibinfo
  {journal} {The European Physical Journal A}\ }\textbf {\bibinfo {volume}
  {55}},\ \bibinfo {pages} {93} (\bibinfo {year} {2019})}\BibitemShut {NoStop}%
\bibitem [{\citenamefont {Bally}\ \emph {et~al.}(2014)\citenamefont {Bally},
  \citenamefont {Avez}, \citenamefont {Bender},\ and\ \citenamefont
  {Heenen}}]{Bally14a}%
  \BibitemOpen
  \bibfield  {author} {\bibinfo {author} {\bibfnamefont {B.}~\bibnamefont
  {Bally}}, \bibinfo {author} {\bibfnamefont {B.}~\bibnamefont {Avez}},
  \bibinfo {author} {\bibfnamefont {M.}~\bibnamefont {Bender}}, \ and\ \bibinfo
  {author} {\bibfnamefont {P.-H.}\ \bibnamefont {Heenen}},\ }\href {\doibase
  10.1103/PhysRevLett.113.162501} {\bibfield  {journal} {\bibinfo  {journal}
  {Phys. Rev. Lett.}\ }\textbf {\bibinfo {volume} {113}},\ \bibinfo {pages}
  {162501} (\bibinfo {year} {2014})}\BibitemShut {NoStop}%
\bibitem [{\citenamefont {Dufour}\ and\ \citenamefont
  {Zuker}(1996)}]{Dufour96a}%
  \BibitemOpen
  \bibfield  {author} {\bibinfo {author} {\bibfnamefont {M.}~\bibnamefont
  {Dufour}}\ and\ \bibinfo {author} {\bibfnamefont {A.~P.}\ \bibnamefont
  {Zuker}},\ }\href {\doibase 10.1103/PhysRevC.54.1641} {\bibfield  {journal}
  {\bibinfo  {journal} {Phys. Rev. C}\ }\textbf {\bibinfo {volume} {54}},\
  \bibinfo {pages} {1641} (\bibinfo {year} {1996})}\BibitemShut {NoStop}%
\bibitem [{\citenamefont {Talmi}(1993)}]{Talmi93a}%
  \BibitemOpen
  \bibfield  {author} {\bibinfo {author} {\bibfnamefont {I.}~\bibnamefont
  {Talmi}},\ }\href@noop {} {\emph {\bibinfo {title} {Simple models of complex
  nuclei}}}\ (\bibinfo  {publisher} {CRC Press},\ \bibinfo {year}
  {1993})\BibitemShut {NoStop}%
\bibitem [{\citenamefont {Hinohara}\ and\ \citenamefont
  {Engel}(2014)}]{Hinohara14a}%
  \BibitemOpen
  \bibfield  {author} {\bibinfo {author} {\bibfnamefont {N.}~\bibnamefont
  {Hinohara}}\ and\ \bibinfo {author} {\bibfnamefont {J.}~\bibnamefont
  {Engel}},\ }\href {\doibase 10.1103/PhysRevC.90.031301} {\bibfield  {journal}
  {\bibinfo  {journal} {Phys. Rev. C}\ }\textbf {\bibinfo {volume} {90}},\
  \bibinfo {pages} {031301} (\bibinfo {year} {2014})}\BibitemShut {NoStop}%
\bibitem [{\citenamefont {Bally}\ \emph {et~al.}(2019)\citenamefont {Bally},
  \citenamefont {S\'anchez},\ and\ \citenamefont {Rodr\'iguez}}]{Bally19a}%
  \BibitemOpen
  \bibfield  {author} {\bibinfo {author} {\bibfnamefont {B.}~\bibnamefont
  {Bally}}, \bibinfo {author} {\bibfnamefont {A.}~\bibnamefont {S\'anchez}}, \
  and\ \bibinfo {author} {\bibfnamefont {T.~R.}\ \bibnamefont {Rodr\'iguez}},\
  }\href@noop {} {\bibfield  {journal} {\bibinfo  {journal} {in preparation}\ }
  (\bibinfo {year} {2019})}\BibitemShut {NoStop}%
\bibitem [{\citenamefont {Vaquero}\ \emph {et~al.}(2011)\citenamefont
  {Vaquero}, \citenamefont {Rodr\'iguez},\ and\ \citenamefont
  {Egido}}]{Lopez11a}%
  \BibitemOpen
  \bibfield  {author} {\bibinfo {author} {\bibfnamefont {N.~L.}\ \bibnamefont
  {Vaquero}}, \bibinfo {author} {\bibfnamefont {T.~R.}\ \bibnamefont
  {Rodr\'iguez}}, \ and\ \bibinfo {author} {\bibfnamefont {J.~L.}\ \bibnamefont
  {Egido}},\ }\href {\doibase https://doi.org/10.1016/j.physletb.2011.09.073}
  {\bibfield  {journal} {\bibinfo  {journal} {Physics Letters B}\ }\textbf
  {\bibinfo {volume} {704}},\ \bibinfo {pages} {520 } (\bibinfo {year}
  {2011})}\BibitemShut {NoStop}%
\bibitem [{\citenamefont {Borrajo}\ \emph {et~al.}(2015)\citenamefont
  {Borrajo}, \citenamefont {Rodr\'iguez},\ and\ \citenamefont
  {Egido}}]{Borrajo15a}%
  \BibitemOpen
  \bibfield  {author} {\bibinfo {author} {\bibfnamefont {M.}~\bibnamefont
  {Borrajo}}, \bibinfo {author} {\bibfnamefont {T.~R.}\ \bibnamefont
  {Rodr\'iguez}}, \ and\ \bibinfo {author} {\bibfnamefont {J.~L.}\ \bibnamefont
  {Egido}},\ }\href {\doibase https://doi.org/10.1016/j.physletb.2015.05.030}
  {\bibfield  {journal} {\bibinfo  {journal} {Physics Letters B}\ }\textbf
  {\bibinfo {volume} {746}},\ \bibinfo {pages} {341 } (\bibinfo {year}
  {2015})}\BibitemShut {NoStop}%
\bibitem [{\citenamefont {Egido}\ \emph {et~al.}(2016)\citenamefont {Egido},
  \citenamefont {Borrajo},\ and\ \citenamefont {Rodr\'{\i}guez}}]{Egido16b}%
  \BibitemOpen
  \bibfield  {author} {\bibinfo {author} {\bibfnamefont {J.~L.}\ \bibnamefont
  {Egido}}, \bibinfo {author} {\bibfnamefont {M.}~\bibnamefont {Borrajo}}, \
  and\ \bibinfo {author} {\bibfnamefont {T.~R.}\ \bibnamefont
  {Rodr\'{\i}guez}},\ }\href {\doibase 10.1103/PhysRevLett.116.052502}
  {\bibfield  {journal} {\bibinfo  {journal} {Phys. Rev. Lett.}\ }\textbf
  {\bibinfo {volume} {116}},\ \bibinfo {pages} {052502} (\bibinfo {year}
  {2016})}\BibitemShut {NoStop}%
\bibitem [{\citenamefont {Rodr\'{\i}guez}\ \emph {et~al.}(2016)\citenamefont
  {Rodr\'{\i}guez}, \citenamefont {Poves},\ and\ \citenamefont
  {Nowacki}}]{Rodriguez16a}%
  \BibitemOpen
  \bibfield  {author} {\bibinfo {author} {\bibfnamefont {T.~R.}\ \bibnamefont
  {Rodr\'{\i}guez}}, \bibinfo {author} {\bibfnamefont {A.}~\bibnamefont
  {Poves}}, \ and\ \bibinfo {author} {\bibfnamefont {F.}~\bibnamefont
  {Nowacki}},\ }\href {\doibase 10.1103/PhysRevC.93.054316} {\bibfield
  {journal} {\bibinfo  {journal} {Phys. Rev. C}\ }\textbf {\bibinfo {volume}
  {93}},\ \bibinfo {pages} {054316} (\bibinfo {year} {2016})}\BibitemShut
  {NoStop}%
\bibitem [{\citenamefont {Poves}\ \emph {et~al.}(2019)\citenamefont {Poves},
  \citenamefont {Nowacki},\ and\ \citenamefont {Alhassid}}]{Poves19a}%
  \BibitemOpen
  \bibfield  {author} {\bibinfo {author} {\bibfnamefont {A.}~\bibnamefont
  {Poves}}, \bibinfo {author} {\bibfnamefont {F.}~\bibnamefont {Nowacki}}, \
  and\ \bibinfo {author} {\bibfnamefont {Y.}~\bibnamefont {Alhassid}},\
  }\href@noop {} {\  (\bibinfo {year} {2019})},\ \Eprint
  {http://arxiv.org/abs/1906.07542} {arXiv:1906.07542 [nucl-th]} \BibitemShut
  {NoStop}%
\bibitem [{\citenamefont {Ripoche}\ \emph {et~al.}(2017)\citenamefont
  {Ripoche}, \citenamefont {Lacroix}, \citenamefont {Gambacurta}, \citenamefont
  {Ebran},\ and\ \citenamefont {Duguet}}]{Ripoche17a}%
  \BibitemOpen
  \bibfield  {author} {\bibinfo {author} {\bibfnamefont {J.}~\bibnamefont
  {Ripoche}}, \bibinfo {author} {\bibfnamefont {D.}~\bibnamefont {Lacroix}},
  \bibinfo {author} {\bibfnamefont {D.}~\bibnamefont {Gambacurta}}, \bibinfo
  {author} {\bibfnamefont {J.-P.}\ \bibnamefont {Ebran}}, \ and\ \bibinfo
  {author} {\bibfnamefont {T.}~\bibnamefont {Duguet}},\ }\href {\doibase
  10.1103/PhysRevC.95.014326} {\bibfield  {journal} {\bibinfo  {journal} {Phys.
  Rev. C}\ }\textbf {\bibinfo {volume} {95}},\ \bibinfo {pages} {014326}
  (\bibinfo {year} {2017})}\BibitemShut {NoStop}%
\bibitem [{\citenamefont {Ripoche}\ \emph {et~al.}(2018)\citenamefont
  {Ripoche}, \citenamefont {Duguet}, \citenamefont {Ebran},\ and\ \citenamefont
  {Lacroix}}]{Ripoche18a}%
  \BibitemOpen
  \bibfield  {author} {\bibinfo {author} {\bibfnamefont {J.}~\bibnamefont
  {Ripoche}}, \bibinfo {author} {\bibfnamefont {T.}~\bibnamefont {Duguet}},
  \bibinfo {author} {\bibfnamefont {J.-P.}\ \bibnamefont {Ebran}}, \ and\
  \bibinfo {author} {\bibfnamefont {D.}~\bibnamefont {Lacroix}},\ }\href
  {\doibase 10.1103/PhysRevC.97.064316} {\bibfield  {journal} {\bibinfo
  {journal} {Phys. Rev. C}\ }\textbf {\bibinfo {volume} {97}},\ \bibinfo
  {pages} {064316} (\bibinfo {year} {2018})}\BibitemShut {NoStop}%
\end{thebibliography}%

\end{document}